\documentclass[prx,longbibliography,showpacs,twocolumn,superscriptaddress,amsmath,amssymb,verbatim]{revtex4-2}
\usepackage{amsmath,amssymb,amsfonts,stmaryrd,wasysym,graphicx,multirow,textcomp,subfigure,makecell}
\usepackage{url}
\usepackage[colorlinks=true,citecolor=blue,urlcolor=blue]{hyperref}
\usepackage{mathtools}
\usepackage[utf8]{inputenc}
\usepackage{braket}
\usepackage{amsmath}
\usepackage{xcolor}
\usepackage{color,soul} 
\usepackage{bm,xfrac}
\usepackage[normalem]{ulem}

\usepackage{enumitem}
\usepackage{dcolumn}
\usepackage{bm}
\usepackage[flushleft]{threeparttable}
\usepackage{braket}
\usepackage{array}
\usepackage{booktabs}
\usepackage{float}
\usepackage{setspace}
\hypersetup{colorlinks=true, urlcolor=blue, citecolor=cyan, pdfborder={0 0 0},}

\hypersetup{colorlinks=true, urlcolor=blue, citecolor=cyan, pdfborder={0 0 0}}

\def\be{\begin{equation}}
	\def\ee{\end{equation}}
\def\bea{\begin{eqnarray}}
	\def\eea{\end{eqnarray}}

\newcommand{\bs}{\boldsymbol}
\DeclareMathAlphabet{\bi}{OML}{cmm}{b}{it}

\def\ba{\begin{aligned}}
\def\ea{\end{aligned}}
\def\be{\begin{equation}}
\def\ee{\end{equation}}
\def\bearr{\begin{eqnarray}}
\def\eearr{\end{eqnarray}}
\def\la{\langle}
\def\ra{\rangle}
\def\l{\left}
\def\r{\right}

\begin{document}
\title{Bulk-boundary correspondence in extended trimer Su-Schrieffer-Heeger model}
\bigskip
\author{Sonu Verma}
\email{sonu.vermaiitk@gmail.com}
\affiliation{Center for Theoretical Physics of Complex Systems, Institute for Basic Science (IBS),
Daejeon 34126, Republic of Korea}
\author{Tarun Kanti Ghosh}
\email{tkghosh@iitk.ac.in}
\affiliation{Department of Physics, Indian Institute of Technology-Kanpur,
Kanpur-208 016, India}
\begin{abstract}

We consider an extended trimer Su-Schrieffer-Heeger (SSH) tight-binding Hamiltonian keeping up to next-nearest-neighbor (NNN) 
hopping terms and on-site potential energy. 
The Bloch Hamiltonian can be expressed in terms of all the eight generators (i.e. Gell-Mann matrices) of 
the SU(3) group. We provide exact analytical expressions of three dispersive 
energy bands and the corresponding eigenstates for any choices of the system parameters. 
The system lacks full chiral symmetry since the energy spectrum is not symmetric around zero,
except at isolated Bloch wavevectors.
We explore parity, time reversal, and certain special chiral symmetries for various system parameters. 
We discuss the bulk-boundary correspondence by numerically computing the Zak phase for all 
the bands and the boundary modes in the open boundary condition.  
There are three different kinds of topological phase transitions, which are classified based on the
gap closing points in the Brillouin zone (BZ) while
tuning the nearest-neighbor (NN) and NNN hopping terms.
We find that quantized changes (in units of $\pi$) in two out of three Zak phases characterize these topological phase transitions. 
We propose another bulk topological invariant, namely 
the {\it sub-lattice winding number}, which also characterizes the topological phase transitions changing
from $ \nu^{\alpha} = 0 \leftrightarrow 2 $ and 
$ \nu^{\alpha} = 0 \leftrightarrow 1 \leftrightarrow 2 $ ($\alpha $: sub-lattice index).
The sub-lattice winding number not only provides a relatively 
simple analytical understanding of topological phases but also successfully establishes bulk-boundary correspondence in the absence of inversion symmetry, which may help in characterizing the bulk-boundary correspondence
of systems without chiral and inversion symmetry. 
\end{abstract}

\maketitle

\section{Introduction}
The geometrical and topological aspects of the Bloch bands
in solid state systems have attracted intense research activities
\cite{Shen,Asboth,Vanderbilt,Fuchs,Oka,Lindner}.  
The symmetry-protected topological (SPT) phases of matter can be understood 
in terms of the interplay between symmetry and topology 
\cite{RMP,RMP1}.
For non-interacting systems, the SPT phases are classified in 
the 10-fold periodic table according to the system's spatial 
dimension and internal symmetries, which include time reversal, 
particle-hole, and chiral symmetries \cite{classi_3D_2008,10fold}. 
The bulk-boundary correspondence is the underlying principle of topological phases 
which relates the non-trivial bulk topological invariant to the number of 
topological boundary modes in the open boundary condition.
Most of the studies are carried out in two-dimensional systems (2D) \cite{Haldane,Kane}, 
however the studies have been extended to one-dimensional (1D) \cite{Chen1,Chen2} and 
three-dimensional (3D) systems \cite{Hasan}  as well.

The simplest model describing the topological features is the dimer
Su-Schrieffer-Heeger (SSH) model \cite{SSH} consists of a 1D 
tight-binding model with alternating hopping parameters.
Further, topological phases of the extended SSH model \cite{gen-SSH}, periodically 
driven 1D optical lattice \cite{1D-op} and quasi-1D lattice \cite{1D,1Da} including the 
next-nearest-neighbour (NNN) hopping have been studied extensively. 
The topological phase of a finite system can be characterized 
by the presence of zero energy boundary modes which are exponentially 
localized at the edges.
The conventional bulk-boundary correspondence using the Zak phase \cite{Zak}, 
similar to the Berry phase \cite{Berry}, has been used as a topological number to classify various 
inversion-symmetric 1D topological systems 
where it is quantized to $\pi$ or 0 (mod $2\pi$) \cite{Delplace}. 
The boundary modes and the Zak phase in various artificial 1D systems such as 
optical lattices \cite{Bloch}, acoustic crystals \cite{Acoustic},  
photonic lattices \cite{Photo-latt}, photonic crystals \cite{Photo-cryst,Photo-cryst1} 
and  artificial spin chains in ultracold atomic system \cite{exp2} 
have been studied elaborately.
However, there is an ambiguity to use the Zak phase as a topological
number, since it depends on the choice of the origin in the real-space 
and how the unit cell is chosen \cite{Bloch,Lee}.
For systems in odd spatial dimensions with chiral and/or inversion symmetry, 
three bulk topological invariants namely the winding number,
the Zak phase, 
and inversion symmetry indicator 
are proposed \cite{Hughes}. 

The momentum space Hamiltonian of 1D, 2D and 3D topological systems 
having two bands is represented by irreducible SU(2) generators.
However, there is a growing interest in three-band in 2D and  
3D topological systems, whose momentum space Hamiltonian 
is represented by an irreducible SU(3) generators \cite{GM,Simon} given in Appendix \ref{gm}. 
For example, Lieb \cite{Lieb,Lieb1,Lieb2} and $\alpha$-$\mathcal{T}_3$ lattices 
\cite{T3,T3a}
in 2D  as well as triple-component fermions \cite{TCF,TCF1,TCF2,TCF3} in 3D exhibits 
three bands, two dispersive bands, and one flat band.
All three bands touch at a given ${\bm k}$ point in the Brillouin zone (BZ), leading to a three-fold
degeneracy at the band touching point. In Ref.~\cite{Pujari_2019}, the geometric phase structure of SU(3) fermions in graphene-like models has also been discussed.
In Ref. \cite{Galitski}, it was proposed that in a square optical lattice 
with spin-orbit coupling subjected to a spatially homogeneous SU(3) 
synthetic gauge field can exhibit nontrivial topology. 
The possibility of designing a
2D lattice model having SU(3) symmetry for spinless fermion has been discussed
in Ref. \cite{Das}.
In a recent study \cite{Ortix},  
a two-dimensional electron gas with Rashba spin-orbit coupling has been
generalized to an SU(3) system, by considering 
trigonal crystal field  effect which leads to
a partial splitting of the energy levels at the time-reversal invariant
momenta.
But there are very few studies \cite{1D,1Da,Trimer-original, Trimer, Zelaya_2024} on multi-band topological systems in 1D.

In this work, we consider a trimer SSH model by considering three sites in a given unit cell. 
Keeping on-site energy and hopping term up to NNN sites, the Bloch Hamiltonian is expressed 
in terms of all the eight generators of the SU(3) group. We obtain exact analytical expressions 
of three dispersive energy bands and the associated eigenstates for arbitrary choices of the system 
parameters. We explore various symmetry properties of the trimer SSH model for various system parameters. 
Since the system does not possess chiral symmetry, we establish the bulk-boundary condition by numerically computing the 
Zak phases ($\phi_\lambda $ with $\lambda =1,2,3 $) for all the bands along with the boundary modes 
in the open boundary condition. Depending on the gap closing points in the BZ, we observe three 
different kinds of topological phase transitions
while tuning NN and NNN hopping terms. We find that quantized changes in two out 
of three Zak phases are necessary to characterize these topological phase transitions, namely 
$ \phi_1 + \phi_3$ changing from $ \phi_1 + \phi_3 = 0 \leftrightarrow 2 \pi $  as well as 
$ \phi_1 + \phi_3 = 0 \leftrightarrow \pi  \leftrightarrow 2\pi$ characterize these topological phase transitions. 
Moreover, we propose another topological invariant, namely the sub-lattice winding number, which also characterizes topological phase transitions similar to the Zak phases. We derive 
exact analytical expressions of the sub-lattice winding number, which changes 
from $ \nu^{\alpha} = 0 \leftrightarrow 2 $ as well as $ \nu^{\alpha} = 0 \leftrightarrow 1 \leftrightarrow 2 $ across the topological phase transitions. We find that the sub-lattice winding number provides a relatively 
simple analytical understanding of topological phases and also successfully establishes the bulk-boundary correspondence without inversion asymmetric systems, which may help characterize the bulk-boundary correspondence  
of systems without chiral and inversion symmetry.

This paper is organized as follows. In Sec. \ref{sec:Hamiltonian}, we describe the trimer SSH model and
obtain the periodic Bloch Hamiltonian dispersive energy bands along with the corresponding eigenstates 
and various symmetries of the system.
In Sec. \ref{sec:bulk-boundary condition}, we discuss the bulk topological invariants Zak phases and sub-lattice winding numbers. In Sec. \ref{sec:bbc1} and \ref{sec:bbc2}, we discuss the bulk-boundary correspondences in the absence and presence of next-nearest neighbor hopping, respectively.
We summarize our results in Sec. \ref{sec:summary}.
In Appendix \ref{gm}, we provide an explicit form of the Gell-Mann matrices. 
We obtain canonical Bloch Hamiltonian and show the equivalence with the
periodic Bloch Hamiltonian in Appendix \ref{app:Hamiltonian}.
The phase diagram characterizing topological phases in  different parameter regimes is given 
in Appendix \ref{Gen-bulk-boundary condition}.
The sub-lattice winding number for the dimer SSH model is discussed in Appendix \ref{App:winding}. The real space profiles of the edge states are discussed in Appendix \ref{App:real_edge}.

\section{SU(3) Hamiltonian and its various properties} \label{sec:Hamiltonian}
\begin{figure}[ht]
\begin{center}
\includegraphics[width=8.0cm]{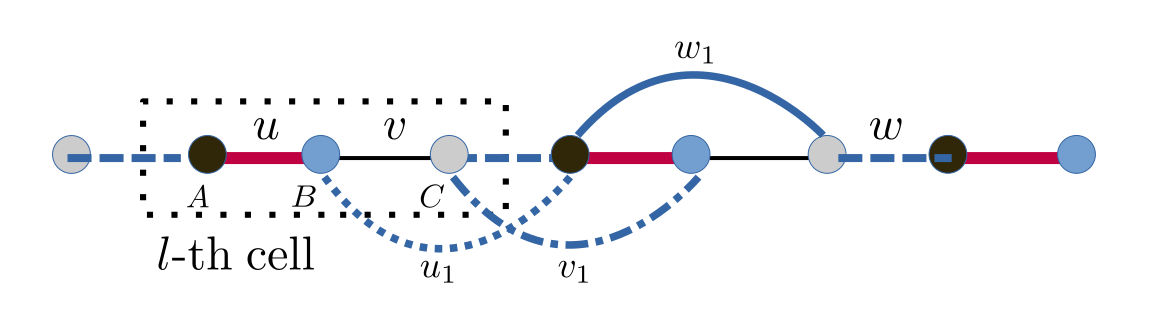}
\end{center}
\caption{Schematic diagram of a trimer SSH model. 
The unit cell of unit length is encircled by the dotted rectangle
containing three sites $A, B$, and $C$.   
Here, $u$, $v$ are the intra-cell NN hopping 
terms and $w$ is the inter-cell hopping term. 
Similarly, $u_1$, $v_1$ are the inter-cell NNN hopping terms and $w$ is
the intra-cell NNN hopping term.}
\label{chain}
\end{figure}

We consider a toy model of a 1D chain with three sites, 
labelled by $A, B$ and $C$, in each unit cell as
shown in Fig. \ref{chain}. 
Considering nearest-neighbor (NN) and NNN hopping terms
along with the on-site interaction energies $\pm \Delta$ at the sites
$A$ and $C$ respectively, 
the trimer SSH tight-binding Hamiltonian in real space reads
\bearr  \label{real-H}
H & = &  \sum_{l} [u a_l^\dagger b_l
+ v b_l^\dagger c_l + w a_{l+1}^\dagger c_{l} + h.c.] \nonumber \\ 
& + & \sum_{l} [ w_1 a_l^\dagger c_l  +
u_1 b_l^\dagger a_{l+1}  +
v_1 c_{l}^\dagger b_{l+1} + h.c.] \nonumber \\
& + & \Delta \sum_{l}[a_l^\dagger a_l -c_l^\dagger c_{l}].
\eearr
Here the operators $ a_l^\dagger (a_l)$, $ b_l^\dagger (b_l)$ and
$ c_l^\dagger (c_l)$  create (destroy) a spinless 
fermion at the sub-lattice sites $A, B$ and $C$ in the $l$-th unit cell of unit
length and $h.c.$ stands for the hermitian conjugate.
As opposed to the dimer SSH model, the NNN hopping occurs between two different sites 
in the same cell as well as in the
next cell.
In the absence of NNN hopping amplitudes and on-site energy,
the above Hamiltonian $H$ reduces to that of in Ref. \cite{Trimer-original,Trimer}. 
We will discuss the Bloch Hamiltonian of the trimer SSH model using two 
different basis states, a periodic Bloch Hamiltonian and the canonical Bloch Hamiltonian 
or simply Bloch Hamiltonian which will be discussed in Appendix \ref{app:Hamiltonian}.
Both representations may be useful and convenient to study certain aspects of the systems.

{\bf Periodic Bloch Hamiltonian}: 
Due to translational invariance, the Hamiltonian can be diagonalized using the
cell index $l$.
The real space field operators at the sub-lattice sites $A, B$ and $C$ in the $l$-th cell 
can be expanded as 
\bearr \label{op-ft}
(a_l, b_l, c_l) & = & \frac{1}{\sqrt{N}} \sum_k e^{i l k } (a_k, b_k, c_k), 
\eearr
where $k$ is the Bloch wave vector restricted within the first BZ 
between $-\pi$ to $\pi$ and $N$ is the number of unit cell. 
There are $3N$ lattice sites in the whole chain.
Using the properties
$ 
\sum_l e^{i l (k - k^\prime) } = N \delta_{k, k^\prime},
$
the Hamiltonian reduces to the form
\bearr 
H 
& = & \sum_k \psi_k^\dagger \mathcal{H}(k) \psi_k,
\eearr
where the spinor $\psi_k^\dagger = (a_k^\dagger,b_k^\dagger,c_k^\dagger )$ and 
the periodic Bloch Hamiltonian $\mathcal{H}(k)$ is given by
\bearr
\mathcal{H}(k) & = & 
\left( \begin{array}{ccc}
\Delta & U &  W \\
U^* & 0 &  V\\
W^* & V^* & -\Delta
\end{array} \right),
\eearr
where the complex elements are $ U  = u + u_1e^{-ik} $, $V= v  + v_1e^{-ik}$ and
$W =  w_1 + w e^{-ik}$.
In real space, the trimer SSH Hamiltonian  is a  $ 3N \times   3N $ matrix,
whereas it becomes a $ 3 \times  3 $ matrix $\mathcal{H}(k)$ for each value of $k$. 
While performing the Fourier transformation of the Hamiltonian, 
only the cell number $l$ is used in the Fourier summation (see Eq. \ref{op-ft}),
without mentioning the actual position of the sublattices within each cell. 
As a result, this Hamiltonian is periodic in $2\pi$: 
$ \mathcal{H} (k+2\pi) = \mathcal{H}(k)$.

The Hamiltonian $\mathcal{H}(k)$ can be expressed in terms of all the 
SU(3) generators, eight Gell-Mann matrices given in Appendix \ref{gm}, as
\bearr
\mathcal{H}(k) 
& = & {\bs \Lambda} \cdot {\bm d(k)}.
\eearr
Here ${\bs \Lambda}$ is a vector composed of the Gell-Mann matrices and
${\bm d(k)}$ is an eight-dimensional real vector arising due to various 
hopping terms up to NNN sites in 1D, whose components are 
$d_1 = u + u_1 \cos(k), d_2 =  u_1 \sin(k), d_3 = \Delta/2,
d_4 = w_1 + w \cos(k), d_5 = w \sin(k),
d_6 =v + v_1 \cos(k), d_7 = v_1 \sin(k)  $ and $d_8 = \sqrt{3} \Delta/2$. 
The length of the eight-dimensional vector ${\bm d(k)}$ is
$ L(k) = \sqrt{ \Delta^2 + |U|^2 + |V|^2+|W|^2 }$.
In the absence of the NNN hopping, $ L(k)$ becomes $k$-independent, which is
in sharp contrast to the dimer SSH model. 

\begin{figure}[ht]
        \begin{center}
              \includegraphics[width=9.0cm]{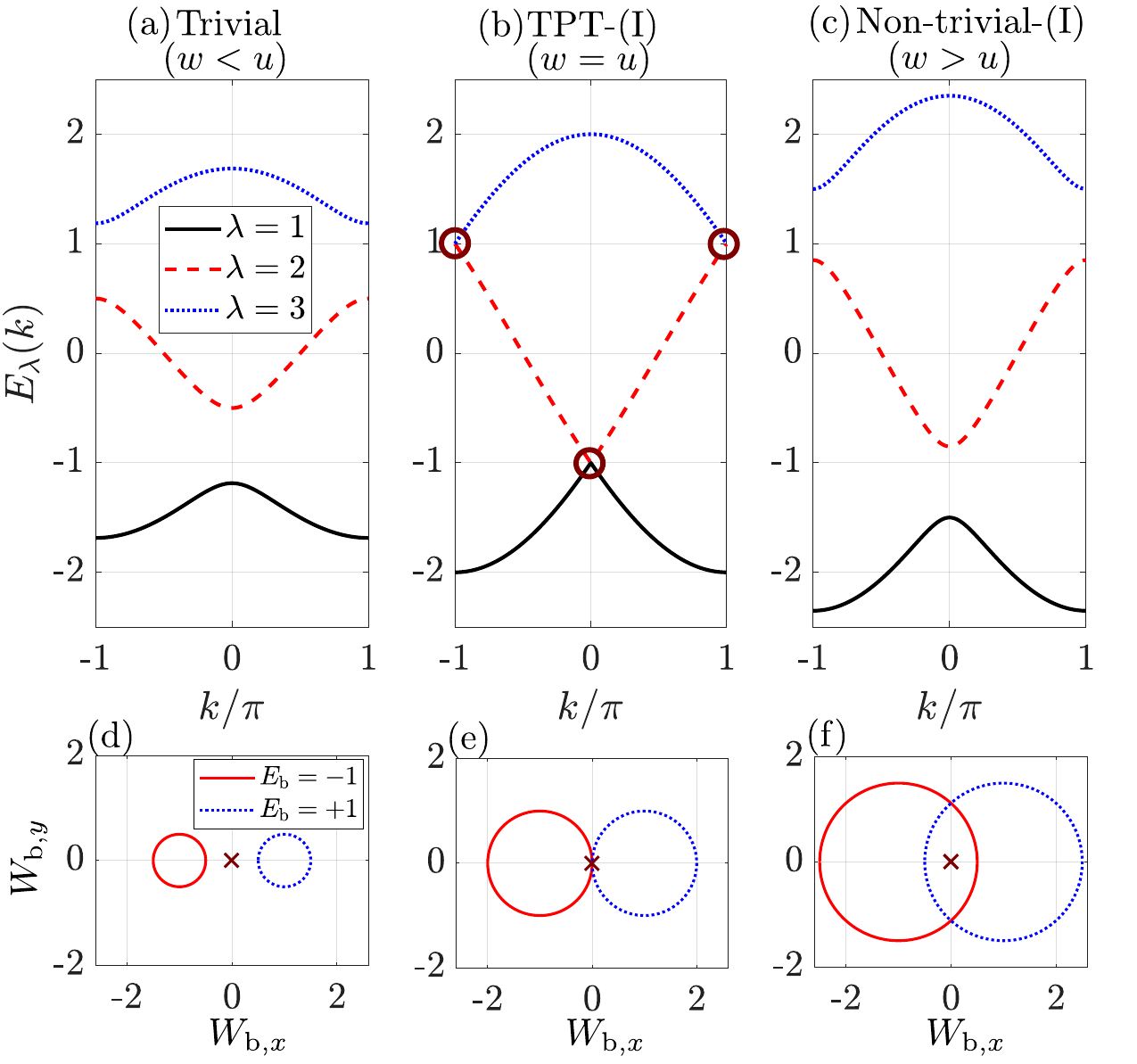}
        \end{center}
        \caption{Upper panel: Dispersion of the three Bloch bands $E_\lambda(k) $ for various
        hopping terms, along with the gap closing points at $k=0,\pm \pi$ in the middle     
        figure. Lower panel: Winding of the complex function 
        $ W_{\text b} \equiv W_{E_{\text b}} = W - E_{\text b}$ is shown in the
        $ W_{{\text b},x}$-$ W_{{\text b},y}$ plane 
        with $ W_{{\text b},x} = {\rm Re}[W_{\text b}]$ and 
        $ W_{{\text b},y} = {\rm Im}[W_{\text b}]$.
        The symbol $\times $ denotes the band
        touching at the origin of the $W_{{\text b},x}$-$ W_{{\text b},y}$ plane, which 
        corresponds to $k=0, \pm \pi$ band touching points in the $E_\lambda $-$k$ plane.
        In both the panels, we have taken $u=v=1$ and $(u_1,v_1,w_1)=(0,0,0)$.}
        \label{fig:E_W_topological phase transitions1}
\end{figure}

\begin{figure*}[htbp!]
		\centering
		\includegraphics[width=1\linewidth]{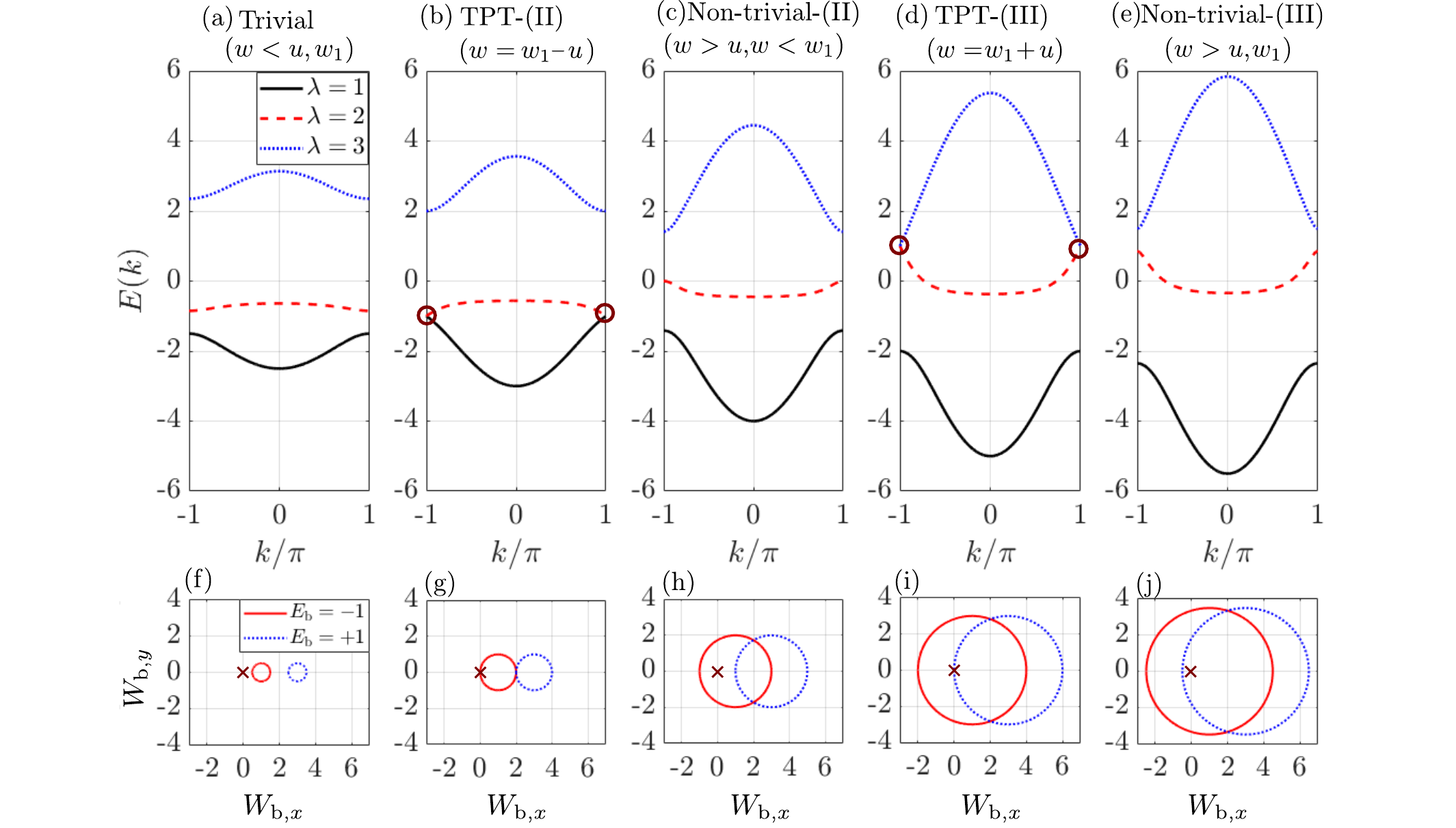}
		\caption{Upper panel: Dispersion of the three Bloch bands $E_\lambda(k) $ for various hopping terms, 
        along with the gap closing points at $k=\pm \pi$ in (b) and (d). Lower panel: Winding of 
        the complex function $ W_{\text b} $ is shown in the $ W_{{\text b},x}$-$ W_{{\text b},y}$ plane.
        In both the panels, we have taken $u=v=1$ and  $(u_1,v_1,w_1)=(0,0,2)$.}
        \label{fig:E_W_topological phase transitions23}
\end{figure*}

\subsection{Energy bands and their properties}
The Hamiltonian, $\mathcal{H}(k)$ yield
the following depressed cubic equation for eigenvalues $E$:
$ 
E^3 + p E + q = 0 
$
with $ p = - L^2(k)$ and 
$q = -[ (|U|^2 - |V|^2 )\Delta + UVW^* + U^*V^*W ]$.
The three energy bands are
\be 
E_\lambda(k) = 2 \sqrt{\frac{-p}{3}} \cos\l[\frac{1}{3} 
\arccos\l(\frac{3q}{2p} \sqrt{\frac{-3}{p}} \r) 
+ \frac{2\pi \lambda}{3} \r]
\ee
with $ \lambda=1,2,3$ and $E_3 \geq E_2 \geq E_1$. 
It is important to note here that the middle band $E_2(k)$ vanishes at two
$k$ points whenever $q=0$, which will be useful in discussing the symmetries of the system.
Therefore the two locations of the zeros of $E_2(k)$ can be found from
the expression of $q$ for the given system parameters. For example, 
$E_2(k = \pm \pi/2) = 0$ for the following two simple cases: 
(i) $\Delta \neq 0 $, $u=v$ and $u_1=v_1=w_1=0$ and (ii)
for any values of  $u,v,w$  with $\Delta = u_1=v_1=w_1=0 $.
Figures \ref{fig:E_W_topological phase transitions1} and \ref{fig:E_W_topological phase transitions23} display the band dispersion for the various system parameters.

The corresponding eigenspinors are obtained as
\be 
\psi_\lambda(k) = N_\lambda
\left( \begin{array}{c}
(E_\lambda + \Delta )U + V^*W   \\
E_\lambda^2 - \Delta^2 - |W|^2  \\
(E_\lambda - \Delta )V^* + UW^* 
\end{array} \right),
\ee
where the $N_\lambda $ is the normalization constant.
It can be easily checked that $ \psi_\lambda(-k) = \psi_\lambda^*(k)$
since $ E_\lambda(-k) = E_\lambda(k) $.

Following Ref. \cite{Simon}, the topology of the energy bands is obtained
as
\bearr
& & E_1(k) + E_2(k) + E_3(k) =  0 \label{plane} \\
& & E_1^2(k) + E_2^2(k) + E_3^2(k)  =  2 L^2(k) \label{sphere}\\
&  & E_3^2(k) - E_1(k)E_2(k) = L^2(k) \\
&  & 2 E_{1,2}(k)  =  - E_3(k) \pm 
\sqrt{ 4 L^2(k)- 3E_3^2(k) }.
\eearr
for any given $k$. In three-dimensional Euclidean space with the axes 
$E_1,E_2, E_3$, Eq. (\ref{plane}) describes a plane passing through 
the origin, for which the normal line has direction cosines: 
$(1/\sqrt{3})(1,1,1) $.
Similarly, Eq. (\ref{sphere}) is a sphere of radius $ \sqrt{2} L(k) $ 
centered at the origin. These properties will be useful in discussing the symmetries of the system.

Here we consider the simplest case when all the NNN
hopping terms are neglected. The parameters $p$ and $q$ are simplified as
$ p = -(\Delta^2 + u^2 +v^2 + w^2)$ and 
$ q = - [\Delta (u^2-v^2) + uvw \cos(k)]$. 
The corresponding eigenvectors in this case are
\be 
\psi_\lambda(k) = N_\lambda
\left( \begin{array}{c}
u (E_\lambda + \Delta ) + v w e^{-ik}  \\
E_\lambda^2 - \Delta^2 - w^2  \\
 v (E_\lambda - \Delta ) + u w e^{ik}
\end{array} \right).
\ee

\subsection{Symmetry properties of the Hamiltonian $\mathcal{H}(k) $ }
Here we will discuss three symmetry properties of the Bloch Hamiltonian $\mathcal{H}(k) $.

{\bf Time-reversal symmetry}:
Note that the three complex Gell-Mann matrices are odd under time reversal:
$\mathcal{C}^{-1} \Lambda_{2,5,7}\mathcal{C} = -\Lambda_{2,5,7} $ and
the remaining matrices are even under time reversal,
with $ \mathcal{C}$ being the complex conjugation operator.
At the same time, the three components of the vector ${\bm d}$ are odd 
under time-reversal: $d_{2,5,7}(-k) = -d_{2,5,7}(k) $ and
the remaining components are even under time reversal.
Thus the Hamiltonian $\mathcal{H}(k) $ is time-reversal invariant:
$ \mathcal{C}\mathcal{H}(-k)\mathcal{C} =  \mathcal{H}^*(-k)=  \mathcal{H}(k) $.
 
{\bf Inversion symmetry}:
The system is invariant 
under space inversion with the inversion point is at the sublattice $B$:
$\mathcal{P} \mathcal{H}(-k) \mathcal{P} =  \mathcal{H}(k)$,
where the unitary parity operator is given by 
$\mathcal{P} = \text{off-diag}(1,1,1)$ with $\mathcal{P} = \mathcal{P}^{-1}$. The inversion operator $ \mathcal{P} $ is equivalent to $\sigma_x$ for the
dimer SSH Hamiltonian. 

{\bf Chiral symmetry}:
The chiral symmetry states that if there
is a state $\psi_k$ with energy $E$ for the given Hamiltonian $H$, 
there  must be another state $ \psi_{k}^\prime = \Gamma \psi_k$ with the
energy $-E$. Here $\Gamma$ is the chiral operator such that 
$ \Gamma H \Gamma^\dagger = - H$. 
Thus the chiral operator relates the positive and negative energies 
and explains the electron-hole symmetry in the bands.

\subsubsection{$\Delta =0$}For $u=v $ and in the presence of NNN hopping $w_1$ and $u_1=v_1$, the space-inversion symmetry is preserved. This is quite different from the dimer SSH model, where the space-inversion symmetry is broken as soon as we consider different NNN hopping terms for different sub-lattices since they involve hopping between
two successive equivalent sites result in a diagonal element in the Bloch
Hamiltonian. 
In the case of the trimer SSH model, hopping between
two successive equivalent sites are only possible when
next to NNN hopping is considered. 
The space-inversion symmetry will be broken 
in the trimer SSH model if we consider the next NNN hopping terms. Moreover, the inversion symmetry is broken for $u\neq v $ and or $u_1\neq v_1$ for any value of $w_1$.

The system lacks full chiral symmetry since the energy bands are not symmetric around zero
for all values of $k$. The concept of point chiral symmetry for $u\neq v$ (inversion asymmetric) has been introduced for $\Delta=u_1=v_1=w_1=0$ case
in Ref. \cite{Trimer,Hidden,Hidden1}.
Mathematically, the point chiral symmetry is stated as
$ \Gamma_p \mathcal{H}(k) \Gamma^\dagger = - \mathcal{H}(k+\pi)$ with
the point chiral operator $\Gamma_p = \text{diag}(1,-1,1)$. It immediately follows
$ E_1(\pi - k) = - E_3(k) $ and $ E_2(\pi - k) = - E_2(k) $. Moreover, we find that the point chiral symmetry is broken in the presence of NNN hopping ($u_1,v_1,w_1\neq 0$).

\subsubsection{$\Delta \neq 0$}
In this case, the inversion symmetry and chiral symmetry are broken for any values of $u,v,w_1,u_1,v_1$.
However, the middle band $E_2$ is zero at $k = \pm \pi/2$ for the two simple
cases: (i) $\Delta \neq 0, u =v, u_1=v_1=w_1 = 0$ and (ii) any $u, v, w$ with 
$\Delta = u_1=v_1=w_1 = 0$. In these two cases, $E_1 = -E_3$ (follows from Eq. \ref{plane}) 
implying the bands are symmetric around zero only at $k = \pm \pi/2$. 
For the case (i), we find an {\it isolated} chiral operator which is
$ \Gamma = \text{off-diag}(1,-1,1)$ and satisfies the two fundamental 
relations: $\Gamma \mathcal{H} \Gamma^\dagger = - \mathcal{H} $ 
and $\psi_3 = \Gamma \psi_1$. 
Similarly, for the case (ii), the isolated chiral operator is
$ \Gamma_a = \mathcal{C} \; \text{diag}(1,-1,1)$ and satisfies the two fundamental 
relations: $\Gamma_a \mathcal{H} \Gamma_a^\dagger = - \mathcal{H} $ 
and $\psi_3 = \Gamma_a \psi_1$.

We extend the concept of point chiral symmetry for $\Delta \neq 0, u_1=v_1=w_1=0 $ along
with $u=v $ case. We find that the point chiral symmetry is preserved even in the
presence of $\Delta $ with the condition $u=v$.  The eigenstates of the shifted particle-hole
symmetry are connected as
$ \Gamma_p \psi_1(\pi - k) = \psi_3(k) $ and 
$  \Gamma_p \psi_2(\pi - k) = \psi_2(k)$.
It should be noted here that
the Bloch phases ($\sim e^{\pm i k}$) arising from the inter-cell and 
intra-cell NNN hopping terms break the point chiral symmetry.

\section{Topological invariants} \label{sec:bulk-boundary condition}
In 1D, no non-trivial topological phase is possible
for systems having no symmetry. For systems in odd spatial dimensions with chiral and/or 
inversion symmetry, three bulk-topological invariants, namely 
the winding number ($\nu$), Zak phase ($\phi_{\lambda}$), and inversion symmetry indicator 
($\chi$) are proposed \cite{Hughes}. These topological invariants can take non-trivial values 
depending on the symmetries of the system and correspond to boundary modes in the open 
boundary condition. The bulk winding number is defined as follows \cite{Ludwig1}

\begin{align}\label{eq:winding_def}
\nu  = 
&\oint_{\rm BZ} \frac{dk}{2\pi i} \partial_{k} \log{(\det{(H(k))})},
\end{align}
where $H(k)$ is the Bloch Hamiltonian.
The winding number can only be quantized and is $\mathbb{Z}$-valued 
for systems respecting chiral symmetries. It is to be noted that the bulk 
winding number for non-chiral system is always zero.

However, for systems having inversion  
symmetry with or without chiral symmetry, the Zak phase can be quantized and 
is $\mathbb{Z}_2$-valued ($0,\pi$ mod $2\pi$).
The Zak phase ($\phi_{\lambda}$) is defined as follows
\begin{align}\label{eq:Zak_def}
\phi_{\lambda} =i \oint_{\rm BZ} dk \langle \psi_{\lambda}(k)|\partial_{k} \psi_{\lambda}(k) \rangle,
\end{align}
where $\psi_{\lambda}(k)$ denotes eigenstate corresponding to the band $\lambda $. 

\begin{figure*}[htbp!]
\begin{center}
\includegraphics[width = 1.0\linewidth]{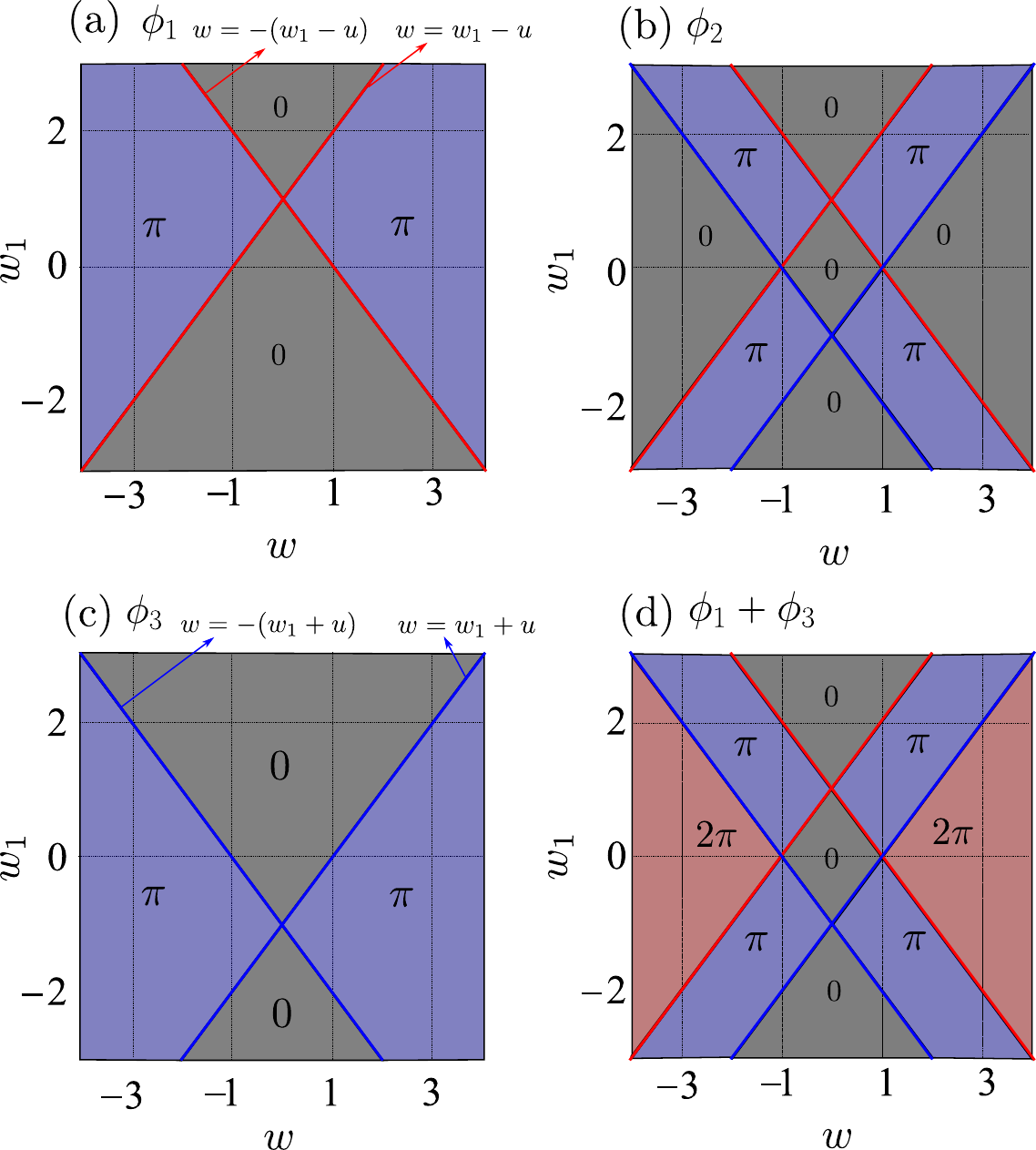}
\end{center}
\caption{\textbf{Bulk-topological invariant}:
(a), (b), and (c) describe the Zak phases ($\phi_1,~\phi_2,~\text{and} ~\phi_3$
in units of $\pi$) for $\lambda=1,~\lambda=2,$ $\lambda=3$ bands, respectively,
in the $(w,w_1)$-plane for $u=v=1$.
(d) The sum of the Zak phases $\phi_1+\phi_3$ (in units of $\pi$) in the $(w,w_1)$-plane for
$u=v=1$. Different lines mark the topological phase transitions. Horizontal dotted lines for $w_1=0$ and $w_1=2.0$ show the parameters of Fig.~\ref{fig:topological phase transitions1_Zak_band_OBC} and Fig.~\ref{fig:topological phase transitions23_Zak_band_OBC}, respectively. }
\label{fig:topological phase transitions3_Zak}
\end{figure*}

\begin{figure*}[htbp!]
\begin{center}
\includegraphics[width = 1.0\linewidth]{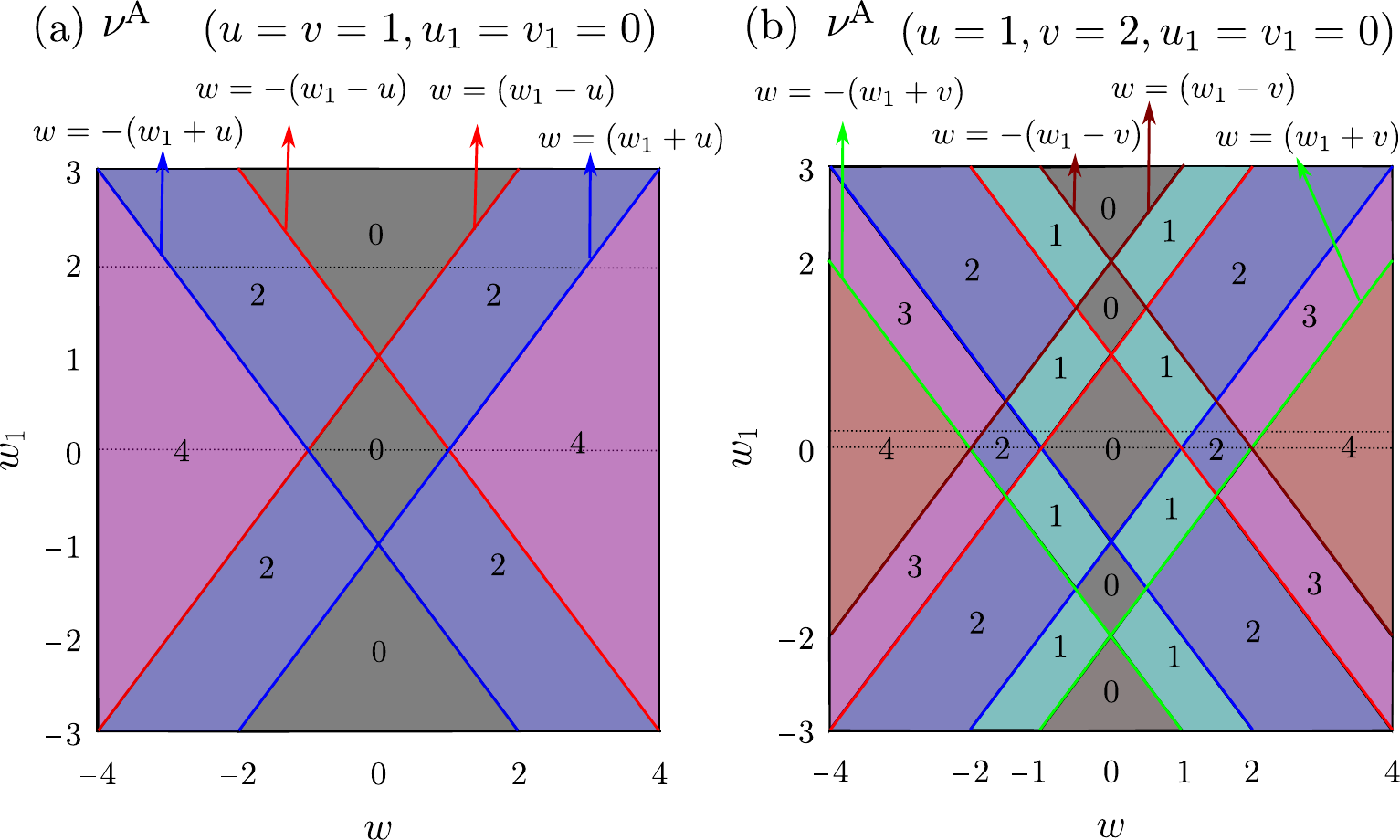}
\end{center}
\caption{\textbf{Bulk-topological invariant: the A sub-lattice winding number}.
(a) and (b) show the behavior of sub-lattice winding number ($\nu_{\rm A}$) in the $(w,w_1)$-plane for the inversion symmetric ($u=v=1, u_1=v_1=0$) and inversion asymmetric ($u=1,v=2, u_1=v_1=0$) chain, respectively. In (a) and (b), different lines and numbers mark the topological phase transition lines and the number of topological edge states that coincide with the sublattice winding number, respectively. In panel (a), horizontal dotted lines for $w_1=0$ and $w_1=2.0$ show the parameters of Fig.~\ref{fig:topological phase transitions1_Zak_band_OBC} and Fig.~\ref{fig:topological phase transitions23_Zak_band_OBC}, respectively. In panel (b), horizontal dotted lines for $w_1=0$ and $w_1=0.2$ show the parameters of Fig.~\ref{fig:topological phase transitions23_winding_band_OBC_w1_0_inv_asym} and Fig.~\ref{fig:topological phase transitions23_winding_band_OBC_w1_finite_inv_asym}, respectively.}
\label{fig:BulkTI_sub_lattice_winding_A}
\end{figure*}

For inversion-symmetric topological insulators with or without chiral symmetry,
the Zak phase has been used to distinguish topologically trivial
($\phi_1=0~\rm mod~ 2\pi)$ and non-trivial phases ($\phi_1=\pi~\rm mod~2\pi$).
However, there is an ambiguity in establishing the bulk-boundary correspondence, as the Zak phase itself depends on the real space origin and choice of unit cell required to define the Bloch Hamiltonian.
In a recent study \cite{Bardarson}, a generalization of
the Zak phase, namely the inter-cellular Zak phase, has been proposed, which establishes
the bulk-boundary correspondence by correctly counting the number of boundary modes
below the Fermi level. The inter-cellular Zak phase 
is the sum of sub-lattice Zak phases \cite{Carpentier} which is defined as follows

\begin{align}\label{eq:sublattice_Zak_def}
\phi_{\lambda}^{\alpha} =i \oint_{\rm BZ} dk
\langle \psi_{\lambda}(k)|P_{\alpha}\partial_{k} P_{\alpha}|\psi_{\lambda}(k) \rangle,
\end{align}
where $P_{\alpha} = |\alpha\ra \la \alpha|  $ 
denotes the projection operator on the sub-lattice $\alpha$ 
with $ |\alpha \ra $ being the sub-lattice eigenstate.


Here we propose a different bulk topological invariant namely the 
\textit{sub-lattice winding number},
similar to the sub-lattice Zak phases defined in Eq.~\ref{eq:sublattice_Zak_def}, 
to discuss the bulk-boundary correspondence in this system. We emphasize here that there are two advantages of defining sub-lattice winding numbers. Firstly, the sub-lattice winding number gives a more straightforward understanding of the topological phases, similar to the winding number for the dimer SSH model~\cite{Asboth}. Secondly, we show that the sub-lattice winding number can still be quantized for the inversion asymmetric chain and successfully establish the bulk-boundary correspondence in the system.

Our system has no non-trivial topological phases as it belongs to class AI of the 10-fold AZ classification of the Hermitian system \cite{classi_3D_2008,10fold}. The point gap-topological phases in non-Hermitian systems also have no non-trivial topological phases according to the 10-fold AZ classification of the Hermitian system. However, the topology of the point-gap non-Hermitian systems can be understood from the point of view of the topology of the doubled Hermitian Hamiltonian [see Refs. \cite{Ueda, Sato} for more details on 38-fold AZ classification of the non-Hermitian system]. For point-gap topological phases, the winding number characterizes the point-gap topology for a given reference eigenvalue $E_{\rm r}$ and Bloch Hamiltonian $\mathcal{H}(k)$, as given by \cite{Ueda, Sato}
\begin{align}
    \nu_{E_{\rm r}}&=\oint_{\rm BZ}\frac{dk}{2\pi i}{\rm Tr}[(\mathcal{H}(k)-E_{\rm r})^{-1}\partial_k(\mathcal{H}(k)-E_{\rm r})]
\end{align}
The non-trivial values of $\nu_{E_{\rm r}}\in\mathbb{Z}$ characterize the point gap topology of the systems and correspond to the edge states (skin modes). This type of non-Hermitian topology has no Hermitian analog and can not be understood from the 10-fold AZ classification of Hermitian systems. However, it can be understood by defining a doubled Hermitian Hamiltonian 
\begin{align}
        \tilde{H}_{E_{\rm r}}(k)=\begin{pmatrix}
        0 & \mathcal{H}(k)-E_{\rm r}\\
        \mathcal{H}^{\dagger}(k)-E_{\rm r}&0
        \end{pmatrix},
\end{align}
which by construction has chiral symmetry for the following chiral 
symmetry operator
\begin{align}
        \Sigma_z=\begin{pmatrix}
        I_{3}& 0\\
        0&-I_{3}
        \end{pmatrix},
 \end{align}
where $I_{3}$ denotes the $3\times3$ identity matrix and belongs to the Class AIII of the 10-fold AZ classification. It is straightforward to see that the winding number $\nu_{E_{\rm r}}$ (winding of the off-diagonal element of $\tilde{\mathcal{H}}_{E_{\rm r}}(k)$) also characterizes the topology of the $\tilde{\mathcal{H}}_{E_{\rm r}}(k)$. Therefore, the skin modes with reference eigenvalue $E_{\rm r}$ correspond to a pair of gapless topological edge modes under the open boundary condition. 

\begin{figure*}[ht]
\begin{center}
\includegraphics[width = 1.0\linewidth]{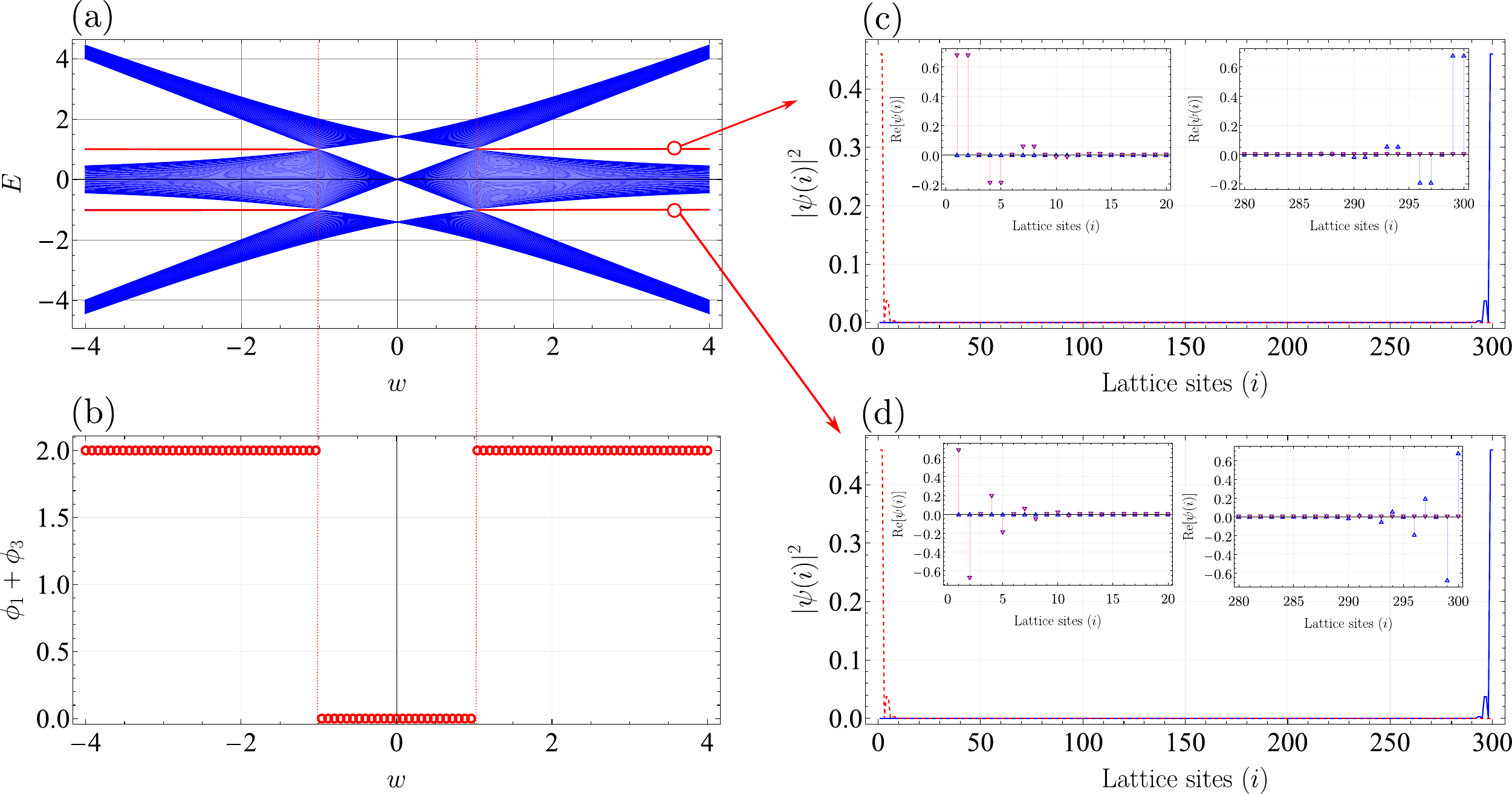}
\end{center}
\caption{\textbf{Bulk-boundary correspondence}:
(a) Energy band structure vs inter-cell hopping amplitude $w$ in the open
boundary condition. The horizontal dashed lines are the edge modes with energy
$E_{\text b} = \pm 1$.
(b) Bulk topological invariant (sum of the Zak phases for the two bands $\phi_1+\phi_3$
(in units of $\pi$) characterizes the appearance of boundary modes and hence the
topological phase transition as one tunes the inter-cell hopping term $w$
for $u=v=1$, and $w_1=0$. 
(c) The real space profile of the wave function amplitude corresponding to the
boundary mode at $E_{\text b}= + 1$ for $(w,w_1)=(3.52,0.0)$.
(d) The real space profile of the probability amplitudes (Inset: wave function amplitudes) corresponding to the
boundary mode at $E_{\text b} = - 1$ for $(w,w_1)=(3.52,0.0)$.}
\label{fig:topological phase transitions1_Zak_band_OBC}
\end{figure*}

\begin{figure*}[ht]
\begin{center}
\includegraphics[width = 1\linewidth]{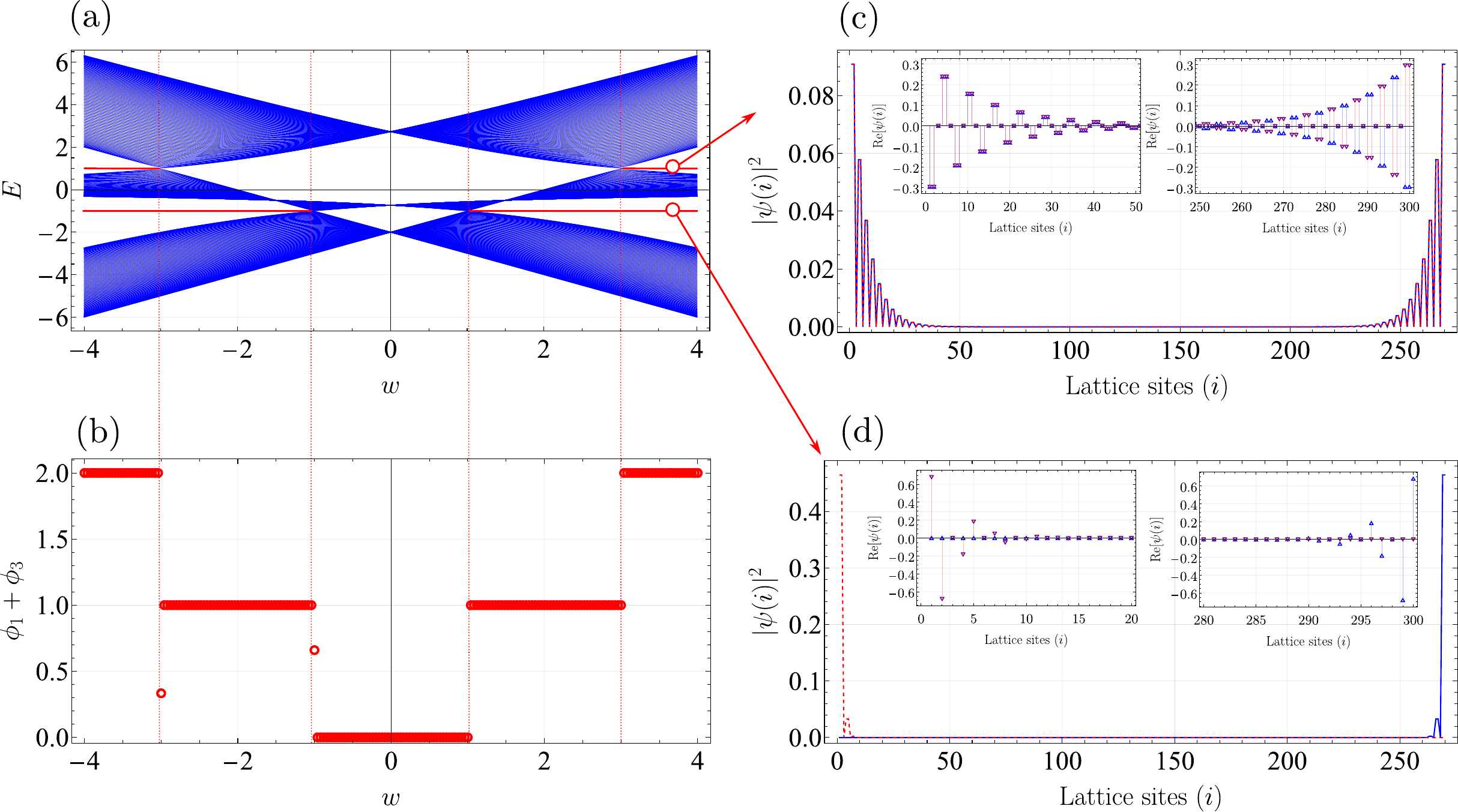}
\end{center}
\caption{\textbf{Bulk-boundary correspondence}:
(a) Energy band structure vs inter-cell hopping term $w$ in the open
boundary condition. The horizontal dashed lines are the edge modes with energy
$E_{\text b} = \pm 1$.
(b) Bulk topological invariant (sum of the Zak phases for the two bands $\phi_1+\phi_3$
(in units of $\pi$) characterizes the appearance of boundary modes and hence the
topological phase transition as one tunes the inter-cell hopping term $w$
for $u=v=1$, and $w_1=2$.
(c) The real space profile of the probability amplitude (Inset: wave function amplitudes) corresponding to the
boundary mode at $E_{\text b} = + 1$ for $(w, w_1)=(3.72, 2.0)$.
(d) The real space profile of the wave function amplitudes corresponding to the boundary
mode at $E_{\text b} = - 1$ for $(w, w_1)=(3.72, 2.0)$.}
\label{fig:topological phase transitions23_Zak_band_OBC}
\end{figure*}

\begin{figure*}[ht]
\begin{center}
\includegraphics[width = 1\linewidth]{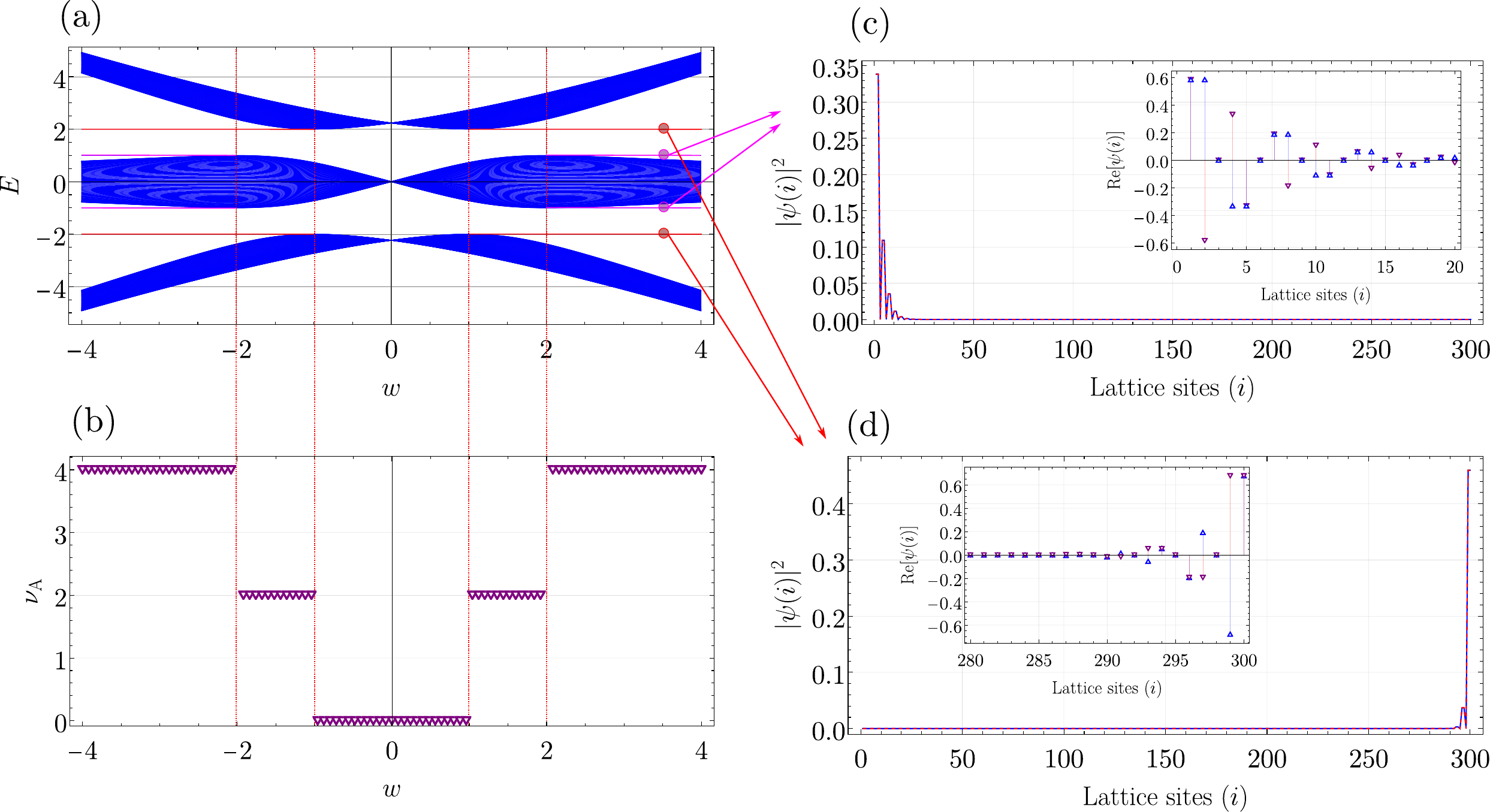}
\end{center}
\caption{\textbf{Bulk-boundary correspondence}:
(a) Energy band structure vs inter-cell hopping term $w$ in the open
boundary condition. The horizontal highlighted lines are the edge modes with energies $E_{\text b} = \pm u$ and $E_{\text b} =\pm v$, which appear at mirror symmetric points $w=\pm u$ and $w=\pm v$, respectively.
(b) Bulk topological invariant (sub-lattice winding number $\nu_{\rm A}$) characterizes the appearance of boundary modes and hence the
topological phase transition as one tunes the inter-cell hopping term $w$
for $u=1,v=2$, and $w_1=u_1=v_1=0$.
(c) The real space profile of the probability amplitudes (Inset: wave function amplitudes) corresponding to the
boundary mode at $E_{\text b} = \pm u$ for $(w, w_1)=(3.52, 0.0)$.
(d) The real space profile of the probability amplitudes (Inset: wave function amplitudes) corresponding to the boundary
mode at $E_{\text b} = \pm v$ for $(w, w_1)=(3.52, 0.0)$.}
\label{fig:topological phase transitions23_winding_band_OBC_w1_0_inv_asym}
\end{figure*}

\begin{figure*}[ht]
\begin{center}
\includegraphics[width = 1\linewidth]{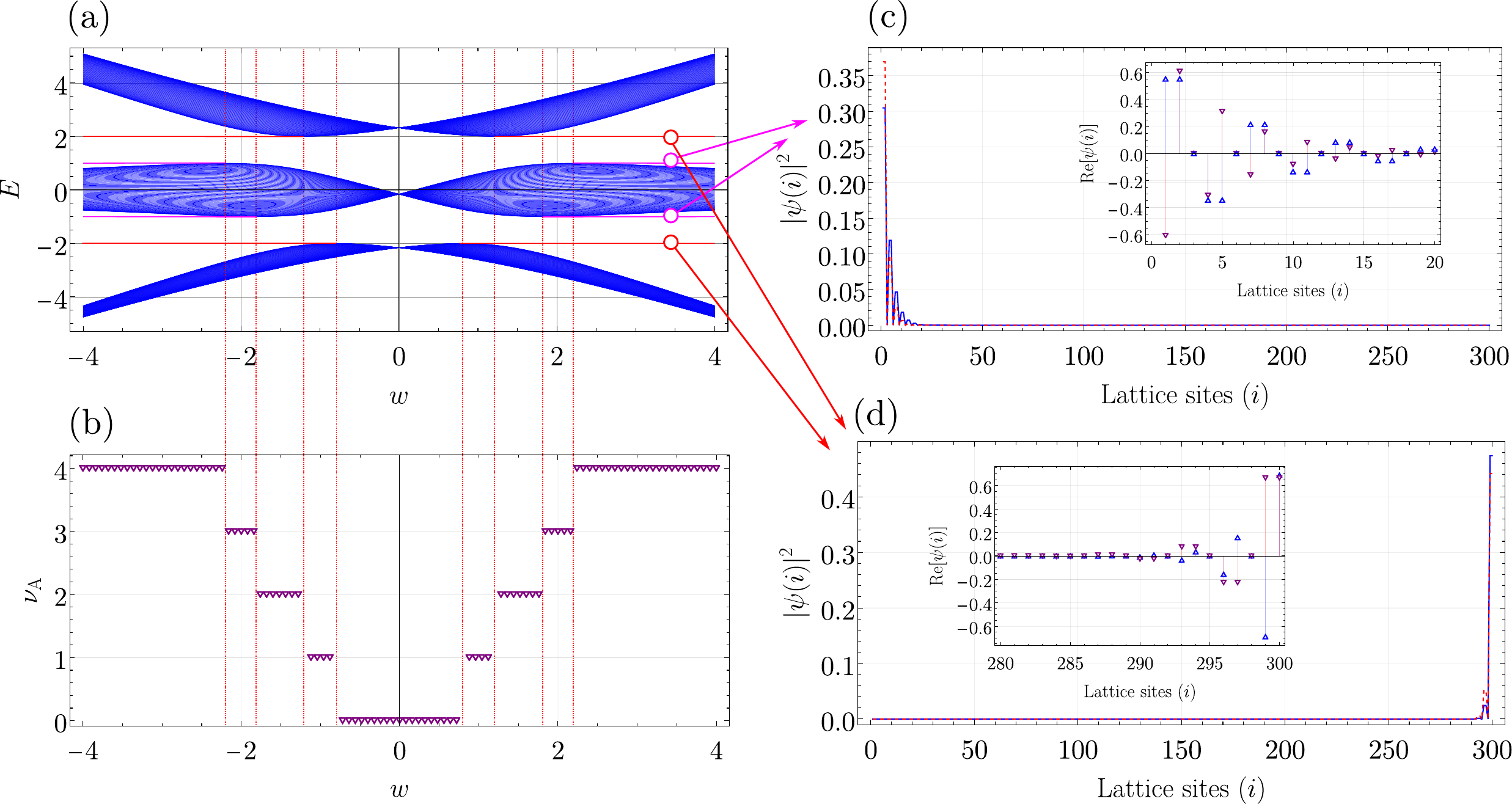}
\end{center}
\caption{\textbf{Bulk-boundary correspondence}:
(a) Energy band structure vs inter-cell hopping term $w$ in the open
boundary condition. The horizontal highlighted lines are the edge modes with energies $E_{\text b} = \pm u$ and $E_{\text b} =\pm v$ which appear at $w=\pm (w_1\pm u)$ and $w=\pm (w_1\pm v)$, respectively.
(b) Bulk topological invariant (sub-lattice winding number $\nu_{\rm A}$) characterizes the appearance of boundary modes and hence the
topological phase transition as one tunes the inter-cell hopping term $w$
for $u=1,v=2$, and $w_1=0.2,u_1=v_1=0$.
(c) The real space profile of the wave function amplitude corresponding to the
boundary mode at $E_{\text b} = \pm u$ for $(w, w_1)=(3.52, 0.2)$.
(d) The real space profile of the wave function amplitude corresponding to the boundary
mode at $E_{\text b} = \pm v$ for $(w, w_1)=(3.52, 0.2)$.}
\label{fig:topological phase transitions23_winding_band_OBC_w1_finite_inv_asym}
\end{figure*}

Following the above arguments, we construct the doubled Hermitian Hamiltonian $\tilde{H}_{E_{\rm b}}(k)$ from the chiral asymmetric Bloch Hamiltonian $H(k)$ via replacing the reference energy $E_{\rm r}$ to the gap-closing energy $E_{\rm b}$. The gaped nature of $\tilde{H}_{E_{\rm b}}(k)$ is defined via $\det(\tilde{H}_{E_{\rm b}}(k))\neq 0$ which is
true for $\det(\mathcal{H}(k)-E_{\rm b})\neq 0$. As our system is Hermitian, it is straightforward to realize that the winding number $\nu_{E_{\rm b}}$ vanishes, as expected. However, we can define the
sub-lattice winding numbers $\nu_{E_{\rm b}}^{\alpha}$ in the following manner,
\begin{align}\label{eq:winding_def_ref}
\nu_{E_{\rm b}}^{\alpha} &= 
\oint_{\rm BZ} \frac{dk}{2\pi i} 
{\rm Tr}[P_{\alpha}(\mathcal{H}(k)-E_{\rm b})^{-1}\partial_{k} (\mathcal{H}(k)-E_{\rm b})P_{\alpha}].
\end{align}
The sub-lattice eigenstate for the trimer SSH model can be expressed
as $|\alpha \ra =(\delta_{\alpha,A}, \delta_{\alpha,B},\delta_{\alpha,C})^{\mathcal T}$ 
with $ {\mathcal T} $ being the transpose operator.
Although the total winding number 
($\nu_{E_{\rm b}}=\sum_{\alpha}\nu_{E_{\rm b}}^{\alpha}=0)$ vanishes
for each reference energy. The sub-lattice winding numbers 
($\nu_{E_{\rm b}}^{\alpha}$) can be quantized and may characterize the number 
of boundary modes in the system. 
This is indeed the case for the present system.
We construct a bulk topological invariant $ \nu^{\rm A} $
which is defined as follows
\begin{align}\label{eq:sub_win_A_bulk_TI}
\nu^{\rm A}=\sum_{E_{\rm b}}\nu_{E_{\rm b}}^{\rm A}.
\end{align}
The non-trivial value of the above topological invariant $\nu^{\rm A}$ counts the number of boundary modes localized at each end of the system and establishes the bulk-boundary correspondence. The physical implication of the zero winding number for the B sub-lattice and equal and opposite winding numbers for the A and C sub-lattices is related to the edge states profile. In the topologically non-trivial phases, the winding number for the A sub-lattice takes finite quantized values that are equal to and opposite to that of the C sub-lattice. The physical consequence is that the edge states have zero wave function amplitudes at the A and C sub-lattice sites for the right and left boundaries of the system, respectively. However, the winding number of the B sub-lattice vanishes, implying the non-vanishing of the wave function amplitudes at both the boundaries of the systems [See Figs.~\ref{fig:topological phase transitions1_Zak_band_OBC}-\ref{fig:topological phase transitions23_winding_band_OBC_w1_finite_inv_asym}]. This is also the case for the dimer SSH model, where the finite winding number for both sub-lattices A and B corresponds to the vanishing of the wave function amplitudes for the boundary modes at the A and B sub-lattice sites for the right and left boundaries of the system, respectively. In the Appendix~\ref{App:real_edge}, we discuss a detailed analysis of the real space profile of the edge states profile.

We note here that the gap closing energy $E_{\rm b}$ does not depend on the size of the system, rather it depends on the system parameters in the Bloch Hamiltonian, \textit{e.g.} $E_{\rm b}=\pm u$, for $u=v$. Therefore, the sublattice winding number $\nu_{\rm A}$ [Eq.~\eqref{eq:sub_win_A_bulk_TI}] can be used to establish the bulk-boundary correspondence.

\begin{widetext}
\begin{center}
\begin{table} 
\def\arraystretch{2.5}
   \begin{tabular}{|c|c|c|c|c|}
   \hline 
  TPT   & System parameters & Bulk band gaps ($\Delta_{\lambda \lambda^\prime}$)  &  \makecell{ Zak phases
  }  & \makecell{Sub-lattice winding number } \\
        \hline
        TPT-I & \makecell{ $ w =|u|, w_1=u_1 = v_1= 0 $ }  & Both $\Delta_{21} $, $\Delta_{32}$ close.  & 
        $ \phi_1 + \phi_3=0  \leftrightarrow  2\pi $ &  $ \nu^A = 0 \leftrightarrow 2 $ \\
        \hline
        TPT-II & \makecell{ $ w =|w_1-u|, w_1, u_1=v_1 \neq 0 $ } & Only $\Delta_{21}$ closes. 
        &$ \phi_1 + \phi_3=0  \leftrightarrow  ~\pi $ & $ \nu^A = 0 \leftrightarrow 1 $\\
        \hline
        TPT-III & \makecell{ $ w =|w_1+u|, w_1, u_1=v_1 \neq 0 $} & Only $\Delta_{32}$ closes.
        & $ \phi_1 + \phi_3=\pi \leftrightarrow  2\pi $  & $ \nu^A = 1 \leftrightarrow 2 $ \\
        \hline
   \end{tabular}
\caption{Classification of topological phase transition (TPT) for inversion symmetric system ($u=v$): Three topological phase transitions for various system parameters based on the band gaps 
   $\Delta_{21}= E_2 - E_1 $ and $ \Delta_{32} = E_3 - E_2 $ closing points along with 
   the topological invariant quantities such as Zak phase and sub-lattice winding number [see Figs. \ref{fig:topological phase transitions3_Zak}, \ref{fig:BulkTI_sub_lattice_winding_A}, and \ref{fig:topologicalphasetransitions123_band}].}
   \label{table:tab}
   \end{table}
       \end{center}
\end{widetext}

\section{Topological phases and bulk-boundary correspondence in the absence
of NNN hopping} \label{sec:bbc1}


Recently, the bulk-boundary correspondence in the trimer SSH model
\cite{Trimer-original,Trimer} without NNN hopping terms 
has been studied. The trimer SSH model lacks full chiral symmetry but 
has generalized chiral, isolated, and point chiral symmetries. Therefore, the chiral winding 
number cannot characterize the topological phases in this system. However, the Zak phase 
successfully establishes the bulk-boundary correspondence for the system parameters respecting the inversion symmetry. 
Moreover, for the trimer SSH model in the absence of NNN hopping term with broken inversion symmetry, 
another generalization of the Zak phase, namely \textit{normalized sub-lattice Zak phase} has been proposed, which 
successfully establishes the bulk-boundary correspondence \cite{Trimer}. In this section, we discuss the bulk-boundary correspondence in the absence of NNN hopping terms ($u_1,~v_1,~w_1$) (Fig.~\ref{chain}).

\subsection{Bulk case}
 {\it $w_1=0$ case}: In the absence of NNN coupling ($w_1=0$) and for $u=v=w$, 
the lower band touches the middle one at $\Gamma$-point ($k=0$) and the 
the middle band touches the upper one at $X$-point
($k=\pi$) simultaneously signaling a topological phase transition 
(See Fig~\ref{fig:E_W_topological phase transitions1}).

We numerically compute the Zak phase defined in Eq.~\ref{eq:Zak_def} for the 
individual bands. Figures~\ref{fig:topological phase transitions3_Zak} (a), (b), (c), and (d) show the 
variation of the Zak phases $\phi_1$, $\phi_2$, $\phi_3$, and 
$\phi_1+\phi_3$ (in units of $\pi$) in the ($w,w_1$)-plane, respectively.
In the absence of NNN hopping ($w_1=0$), our results are in full 
agreement with that of Ref. \cite{Trimer-original,Trimer}. 
It can be seen that any two of the three Zak phases change in a quantized manner 
whenever the bulk band gap closes, which signals a topological phase transition.
Accordingly, in Table \ref{table:tab},
we label all three different topological phase transitions
characterized by the locations of the gap-closing points in the BZ
while tuning $w$ and $w_1$.

To elaborate on the above discussion, we again consider the trimer SSH chain with $u_1=v_1=0,~u=v=1$.
In this case, the bulk energy band gap closes at $\pm1$ therefore we choose the reference
energies $E_{\rm b}=\pm1$.  The sub-lattice winding number $\nu_{\pm 1}^{\rm A}$ take the following form
 \begin{align}
 \nu_{\pm 1}^{\rm A}&=\int_{-\pi} ^{\pi} \partial_{k}\log(W^{*}\pm 1)\frac{dk}{2\pi i}\nonumber\\
 &=\begin{cases}
 0,~&|w_1\pm 1|>w\\
 1,~&|w_1\pm 1|<w,
 \end{cases}
 \end{align}
where $W=w_1+w e^{-i k}$ is nothing but the off-diagonal element of
$\mathcal{H}(k)$. The sub-lattice winding number $\nu_{\pm 1}^{\rm B}$ vanishes and 
$\nu_{\pm 1}^{\rm C}=-\nu_{\pm 1}^{\rm A}$ due to 
$\sum_{\alpha}\nu_{\pm 1}^{\alpha}=0$. It is straightforward to verify from
Fig.~\ref{fig:topological phase transitions1_Zak_band_OBC} and Fig.~\ref{fig:topological phase transitions23_Zak_band_OBC}
that $\nu^{\rm A}=\nu_{+1}^{\rm A}+\nu_{-1}^{\rm A}$ correctly counts the number of
boundary modes at each end of the system. 
For example, in the case of $u_1=v_1=w_1=0$,
for $w>u=v$ and $\nu^{\rm A}=2$ determines one boundary mode with energy $ E_{\rm b} = -1$ and one boundary
mode with energy $ E_{\rm b} = + 1$ at each end of the system. 

One of the advantages of this topological invariant is its analytical 
form and simple understanding of the topological phase transition,
similar to the dimer SSH model \cite{Delplace}.
When $k$ varies across the first BZ, the endpoint of the vector 
${\bf W}_{\text b} = W-E_{\text b}\equiv\hat i W_{{\text b},x} + \hat j W_{{\text b},y} $, with $W=w_1+w e^{-i k},~W_{{\text b},x}={\rm Re}[W]-E_{\rm b},~W_{{\rm b},y}={\rm Im}[W]$,
traces out a circle of radius $ w $
centered at $(w_1 - E_{\text b})$ in the $W_{{\text b},x}$-$W_{{\text b},y}$ plane, 
as shown in the lower panel of Fig. \ref{fig:E_W_topological phase transitions1}.
The gap closes at ${\bf W}_{\text b} =0$, which corresponds to the gap closing points
$k=0,\pm \pi$ in $k$ space. 
The sub-lattice winding number will be one when the trajectory of 
${\bf W}_{\text b}$ encloses the origin, otherwise zero.  
It is clear from Fig.~\ref{fig:E_W_topological phase transitions1} (d), (e), and (f) that 
the winding of the complex function $W_{\text b}$ is zero, ill-defined,
and two for the trivial, critical, and non-trivial phases. 
The topological phase transition
takes place at $w=u=v$ where the contour of $W_{\rm b}$ touches the origin which corresponds
to the bulk-gap closing points at $k=0,\pi$.
Figure  \ref{fig:E_W_topological phase transitions1} displays that tuning the parameter $w$ while fixing $u=v=1$ and
 $(u_1,v_1,w_1)=(0,0,0)$ induces a topological phase transition
characterized by the sub-lattice winding number from $\nu^A = 0 \leftrightarrow 2$. Figure~\ref{fig:BulkTI_sub_lattice_winding_A} (a) shows the variation of the sublattice winding number in the $(w,w_1)$-plane, which shows that non-trivial values of the sublattice winding number coincide with the number of boundary modes in the open boundary condition.

{\it Inversion asymmetric case:}
For the inversion asymmetric chain with $u\neq v,~w_1=u_1=v_1=0$, the boundary modes do not appear at the band-gap closing point; instead, they appear at the mirror-symmetric points at $w=\pm u$ and $w=\pm v$ with energies $E_{\rm b}=\pm u$ and $E_{\rm b}=\pm v$, respectively~\cite{Trimer}. In this case, the bulk-boundary correspondence has been discussed in detail in Ref.~\cite{Trimer} using the normalized sub-lattice Zak phase as a bulk topological invariant determined under the open boundary condition.
\\
In this case, we also find that the sub-lattice winding number can still be quantized and successfully establish the bulk-boundary correspondence. Figure~\ref{fig:BulkTI_sub_lattice_winding_A} (b) shows the variation of the sublattice winding number in the $(w,w_1)$-plane. The non-trivial values of the sublattice winding number coincide with the number of boundary modes in the open boundary condition. To elaborate 
 on the above findings, we show the energy band structure of a finite chain vs $w$, the sublattice winding number vs $w$, and the real space profile of the edge states in Fig.~\ref{fig:topological phase transitions23_winding_band_OBC_w1_0_inv_asym} (a), (b), and [(c)-(d)], respectively.  We note here that the sublattice winding number can still be quantized and used to establish the bulk-boundary correspondence for $\Delta\neq 0$ with a more complicated phase diagram, which we do not discuss here.

\subsection{Finite chain}
For $w>u=v$ and $w_1 =0$, the Zak phases $\phi_1, \phi_3$ are quantized to 
$\pi$ while $\phi_2$ vanishes. In the open boundary condition, 
there exists at each energy $E_{\text b}= \pm 1$, a single boundary 
mode localized at each end of the system [see Fig. \ref{fig:topological phase transitions1_Zak_band_OBC} (a) ]. 
These results establish the bulk-boundary correspondence.
In this case, the localization lengths for the boundary modes at energies 
$ E_{\text b} = \pm 1$ are the same and are determined by the bulk energy gaps
$ \Delta_{21} $ and $\Delta_{32} $.
 
Figure~\ref{fig:topological phase transitions1_Zak_band_OBC} (a) and (b) show the variation of the band structure 
in the open boundary condition and $\phi_1 +\phi_3$ with respect to $w$, respectively.
Figures ~\ref{fig:topological phase transitions1_Zak_band_OBC} (c) and (d) show the real space profile 
of the wave function amplitudes for a finite open chain corresponding to 
the boundary modes with energies $E_{\text b} = +1 $ and  
$E_{\text b} = -1 $, respectively.

In Ref.~\cite{Trimer}, it was shown that the following real space profile of edge states holds due to point chiral symmetry. In the inversion symmetric case, the wave function amplitudes vanish for all edge modes at one of three sub-lattice sites (at the A sub-lattice on the left and C sub-lattice on the right boundary). For the edge states with energies $E_{\rm b}=-u$ ($E_{\rm b}=u$), wave function amplitudes at the A and B sub-lattice sites are equal (equal and opposite). For the edge states with energies $E_{\rm b}=-v$ ($E_{\rm b}=v$), wave function amplitudes at the B and C sub-lattice sites are equal (equal and opposite). 
In the {\it inversion asymmetric} case, the boundary modes lose their inversion partner modes Figs.~\ref{fig:topological phase transitions23_winding_band_OBC_w1_0_inv_asym} (c) and (d). The boundary modes with energies $E_{\rm b}=\pm u$ and $E_{\rm b}=\pm v$ are localized on the left and right boundaries of the system, respectively. However amplitudes of the wave function for $E_{\rm b}=+u$ and $E_{\rm b}=- u$ (similarly for $E_{\rm b}=+v$ and $E_{\rm b}=- v$) do not coincide.

\section{Topological phases and bulk-boundary correspondence in the presence
of NNN hopping} \label{sec:bbc2}
In this section, we discuss the bulk-boundary correspondence in the presence of NNN hopping terms ($u_1,~v_1,~w_1$) (Fig.~\ref{chain}).
\subsection{Bulk}
For finite NNN hopping ($w_1\neq0$), the 
inversion symmetry enforces the hopping to satisfy the following criteria,
\begin{align}
        \Delta= 0 , u=v, ~ \text{and} ~ u_1=v_1.
\end{align}
The results for different sets of $u_1=v_1 \neq 0$
is given in the Appendix \ref{Gen-bulk-boundary condition}. In this case, 
the bulk band structure evolves according to Fig.~\ref{fig:E_W_topological phase transitions23} as 
we tune the inter-cell hopping $w$ for the fixed value of NNN hopping $w_1=2$. 
At $w = u=v $ and $w<w_1$, the band gap between the lower and
middle bands closes at the $X$-point (See Fig. ~\ref{fig:E_W_topological phase transitions23} (b)). 
Similarly, at $w > u=v $ and $w=w_1$, the band gap between the 
middle and the upper bands closes at the $X$-point (See Fig. ~\ref{fig:E_W_topological phase transitions23} (d). We note here that in the presence of NNN hopping the only one of the band gaps 
 $\Delta_{12}$ and $\Delta_{23}$ vanish either at $\Gamma$ or $X-$ points, which signals TPT-II and TPT-III, respectively (see Fig.~\ref{fig:topologicalphasetransitions123_band} in the Appendix.

In the presence of the NNN hopping $(w_1 \neq 0)$, the quantized change
in the Zak phase $ \phi_2$ senses both the gap closings ($\Delta_{21} $ and $ \Delta_{32} $  )  
at $X$ points and  characterizes both the topological phase transitions (See Fig.~\ref{fig:topological phase transitions3_Zak} (b)).
However, the quantized change in the Zak phase  $ \phi_1 $ ($\phi_3 $) senses the band touching 
 $  \Delta_{21} = 0 $ ( $ \Delta_{32} = 0 $) at the $X$-points (See Fig.~\ref{fig:topological phase transitions3_Zak}
(a) and (c)). Hence, the Zak phases $\phi_1$ and $\phi_3$ separately 
characterize the topological phase transition-II (TPT-II) and 
topological phase transition-III (TPT-III), respectively. 

We observe that the addition of 
Zak phases $\phi_1 +\phi_3$ characterize the topological phase transitions 
and number of boundary modes at each end of the system. We note here that the quantized changes in the individual Zak phases $\phi_1$ and $\phi_3$ characterize the topological phase transitions (TPT)-(I and II) and (I and III), respectively. Instead, we use $\phi_1 +\phi_3$ to convey the same information.
A similar analysis holds for $u_1=v_1\neq 0$. However, the 
phase diagram looks more complex, which is given in Appendix 
\ref{Gen-bulk-boundary condition} (See Fig.~\ref{fig:topological phase transitions23_Zak}).

In this case, we also find that the non-trivial values of the sublattice winding number coincide with the number of boundary modes in the open boundary condition, successfully establishing the BBC in the presence of NNN hopping. In the case of $u_1=v_1 $, for $w>u=v,~w<w_1$,
$\nu^{\rm A}=1$ [Eq.~\ref{eq:sub_win_A_bulk_TI}] determines the one boundary mode with energy $ E_{\rm b} =  -1$ at each end of the system 
and for $w>u=v,~w>w_1$, $\nu^{\rm A}=2$ determines one boundary mode with energy $ E_{\rm b} = -1$ and one boundary
mode with energy $ E_{\rm b} =  +1$ at each end of the system.

It is clear from Fig.~\ref{fig:E_W_topological phase transitions23} (f)-(j) that the winding 
of $W_{\text b}$ is zero, ill-defined, one, one, and two, respectively.
The topological phase transitions take place at $w=u=v,w<w_1$ where contour of $W+1$ 
touches the origin, which corresponds to the bulk-gap closing at $k=\pm \pi$ and  
also at $w=u=v,w=w_1$ where the contour of $W-1$ touches the origin
which corresponds to the bulk-gap closing at $k=\pm \pi$.
Figure  \ref{fig:E_W_topological phase transitions23} displays  a series of topological phase transitions
characterized by the sub-lattice winding number from $\nu^A = 0 \leftrightarrow 1 \leftrightarrow 2$,
while tuning the parameter $w$ for fixed $u=v=1$ and
 $(u_1,v_1,w_1)=(0,0, 2)$. Fig.~\ref{fig:BulkTI_sub_lattice_winding_A} (a) shows the variation of the sublattice winding number in the $(w,w_1)$-plane. To elaborate 
 on the above findings, we show the energy band structure of a finite chain vs $w$, the sublattice winding number vs $w$, and the real space profile of the edge states in Fig.~\ref{fig:topological phase transitions23_Zak_band_OBC} (a), (b), and [(c)-(d)].

 {\it Inversion asymmetric case:} Similar to the case of no NNN hopping, in the inversion asymmetric case ($u\neq v$), we find that the sub-lattice winding number can still be quantized and successfully establishes the BBC in the presence of NNN hopping. For $w_1\neq 0$, the boundary modes appear at $w=\pm(w_1\pm u)$ and $w=\pm(w_1\pm v)$; however, their energies $E_{\rm b}=\pm u$ and $E_{\rm b}=\pm v$ (which remain the same), respectively. Fig.~\ref{fig:BulkTI_sub_lattice_winding_A} (b) shows the variation of the sublattice winding number in the $(w,w_1)$-plane. The non-trivial values of the sublattice winding number coincide with the number of boundary modes in the open boundary condition. To elaborate 
 on the above findings, we show the energy band structure of a finite chain vs $w$, the sublattice winding number vs $w$, and the real space profile of the edge states in Fig.~\ref{fig:topological phase transitions23_winding_band_OBC_w1_finite_inv_asym} (a), (b), and [(c)-(d)].

\subsection{Finite chain}
Figs.~\ref{fig:topological phase transitions23_Zak_band_OBC} (a) and (b) show
the variation of the band structure in the open boundary condition and 
$\phi_1 +\phi_3$ with respect
to $w$, respectively.
Figs.~\ref{fig:topological phase transitions23_Zak_band_OBC} (c) and (d) show 
the real space profile of wave function amplitudes for a finite open chain 
corresponding to the boundary modes with energies $E_{\text b} = +1 $  
and $E_{\text b} = -1 $, respectively. Accordingly, in the open boundary condition, 
there exist at each energy $E_{\text b} = \pm 1$, a single boundary 
mode localized at each end of the system. In this case the localization lengths
for the boundary modes at energies $E_{\text b} = \pm 1$ are different and 
are determined by the bulk energy gap
$ \Delta_{21} $ ($ \Delta_{32} $) for negative (positive) energy boundary modes.

In the inversion symmetric case [Figs.~\ref{fig:topological phase transitions23_Zak_band_OBC} (c) and (d)], for all edge modes, the wave function amplitudes vanish at one of three sub-lattice sites (at A sub-lattice on the left and C sub-lattice on the right boundary). For the edge states with energies $E_{\rm b}=-u$ ($E_{\rm b}=u$), wave function amplitudes at the A and B sub-lattice sites are equal (equal and opposite). For the edge states with energies $E_{\rm b}=-v$ ($E_{\rm b}=v$), wave function amplitudes at the B and C sub-lattice sites are equal (equal and opposite).

{\it Inversion asymmetric case:} In this case, the boundary modes lose their inversion partner modes Figs.~\ref{fig:topological phase transitions23_winding_band_OBC_w1_finite_inv_asym} (c) and (d). The boundary modes with energies $E_{\rm b}=\pm u$ and $E_{\rm b}=\pm v$ are localized on the left and right boundaries of the system, respectively. However amplitudes of the wave function for $E_{\rm b}=+u$ and $E_{\rm b}=- u$ (similarly for $E_{\rm b}=+v$ and $E_{\rm b}=- v$) do not coincide. In Ref.~\cite{Trimer}, it was also argued that these real space profile of edge states holds due to point chiral symmetry, which is broken for $w_1\neq 0$). It is unclear which symmetry enforces the properties of the edge states and quantization of the sublattice winding number even in the absence of inversion symmetry ($u\ne v$) and point chiral symmetry ($w_1\neq 0$). To address these queries further, the studies \cite{sonu1,sonu2} of the bulk-boundary correspondences in systems with full or partial symmetry-breaking perturbations using non-Bloch band theory may be useful.

\section{Summary and Conclusion} \label{sec:summary}
In this work, it is shown that the Bloch Hamiltonian of the trimer SSH model with 
the on-site potential energy and up to NNN 
hopping terms can be expressed in terms of all eight generators of the SU(3) group. 
The exact analytical expressions of the three
dispersive energy bands and the associated eigenstates for arbitrary choices of the system parameters are provided.  
We discussed inversion symmetry, time-reversal symmetry, and some special chiral symmetry of the Hamiltonian.
The system lacks full chiral symmetry, however, we find 
three different kinds of topological phase transitions while tuning the NN and NNN hopping terms. 
The topological phase transitions are characterized by the two bulk topological invariants, namely the 
Zak phase and the sub-lattice winding number.
We established the bulk-boundary correspondence by computing the Zak phases ($\phi_\lambda )$ for all the bands along with the boundary modes 
in the open boundary condition. 
The quantized change in two out of three Zak phases  
$ \phi_1 + \phi_3$ changing from $ \phi_1 + \phi_3 = 0 \leftrightarrow 2 \pi $  as well as 
$ \phi_1 + \phi_3 = 0 \leftrightarrow \pi  \leftrightarrow 2\pi$ characterize these topological phase transitions. 
We proposed a new topological invariant, the sub-lattice winding number 
which also characterizes the topological phase transitions. 
The exact analytical expressions of the sub-lattice winding number revealed that 
$\nu^{\alpha} $ changes from $ \nu^{\alpha} = 0 \leftrightarrow 2 $ as well as 
$ \nu^{\alpha} = 0 \leftrightarrow 1 \leftrightarrow 2 $ across the topological phase transitions.
The simple analytical method of calculating the sub-lattice winding number and its success in establishing the bulk-boundary correspondence in the inversion asymmetric system may help understand the bulk-boundary correspondence in systems without chiral and inversion symmetry.

\begin{center}
{\bf ACKNOWLEDGEMENT}
\end{center}
We would like to thank Bashab Dey and SK Firoz Islam for the useful discussion. S.V. acknowledges financial support from the Institute for Basic Science in the Republic of Korea through the project IBS-R024-D1.

\appendix

\section{Gell-Mann matrices and a couple of basic properties} \label{gm}
In this Appendix, we will provide well-known Gell-Mann matrices \cite{GM} and
some of their basic properties.  
The Gell-Mann matrices are
\bearr 
\Lambda_1 & = & \left( 
\begin{array}{ccc}
0 & 1 &  0 \\
1 & 0 &  0 \\
0 & 0 & 0
\end{array} \right), \;
\Lambda_2 = \left( 
\begin{array}{ccc}
0 & -i &  0 \\
i & 0 &  0 \\
0 & 0 & 0
\end{array} \right), \nonumber\\
\Lambda_3 & = & \left( 
\begin{array}{ccc}
1 & 0 &  0 \\
0 & -1 &  0 \\
0 & 0 & 0
\end{array} \right), \; 
\Lambda_4  =  \left( 
\begin{array}{ccc}
0 & 0 &  1 \\
0 & 0 &  0 \\
1 & 0 & 0
\end{array} \right), \nonumber \\ 
\Lambda_5  & = &  \left( 
\begin{array}{ccc}
0 & 0 &  -i \\
0 & 0 &  0 \\
i & 0 & 0
\end{array} \right), \; 
\Lambda_6  =  \left( 
\begin{array}{ccc}
0 & 0 &  0 \\
0 & 0 &  1 \\
0 & 1 & 0
\end{array} \right), \nonumber\\
\Lambda_7 & = &\left( 
\begin{array}{ccc}
0 & 0 &  0 \\
0 & 0 &  -i \\
0 & i & 0
\end{array} \right),
\Lambda_8 = 
\frac{1}{\sqrt{3}}
\left( 
\begin{array}{ccc}
1 & 0 &  0 \\
0 &  1 &  0 \\
0 & 0 & -2
\end{array} \right).
\eearr
These matrices satisfy the Hilbert-Schmidt orthonormality condition
$ {\rm Tr}(\Lambda_{\mu}^\dagger \Lambda_\nu ) = 2 \delta_{\mu \nu} $. 

The particles in the chain are spinless, and the time-reversal operator
in the sub-lattice space is simply the complex conjugation operator $\mathcal{C}$.
The three complex Gell-Mann matrices are odd under time reversal:
$\mathcal{C}^{-1} \Lambda_{2,5,7}\mathcal{C} = -\Lambda_{2,5,7} $ and
the remaining matrices are even under time reversal.

\section{Canonical Bloch Hamiltonian} \label{app:Hamiltonian}
In this Appendix, we will discuss the  canonical Bloch Hamiltonian, 
equivalent to the periodic Bloch Hamiltonian.  
The canonical Bloch Hamiltonian can be obtained by using the following Fourier transformation
\bearr
(a_l, b_l, c_l) & = & \frac{1}{\sqrt{N}} \sum_k e^{i l k  } 
(a_k, e^{i k /3} b_k, e^{i 2k/3} c_k). \nonumber 
\eearr
Here $a_l$, $b_l$ and $c_l$ are the real space field operators at 
the sub-lattice sites $A, B$ and $C$ 
in the $l$-th cell. Note that here exact locations of the sublattices within a cell $l$
are used while taking the Fourier transforms of the field operators.

The real space Hamiltonian $H$ reduces to the form
\bearr 
H 
& = & \sum_k \psi_k^\dagger H(k) \psi_k,
\eearr
where the spinor $\psi_k^\dagger = (a_k^\dagger,b_k^\dagger,c_k^\dagger )$ and 
the canonical Bloch Hamiltonian $H(k)$ is given by
\bearr
H(k) & = & 
\left( \begin{array}{ccc}
\Delta &  e^{ik/3} U &  e^{2ik/3} W \\
  e^{-ik/3} U^* & 0 &   e^{ik/3} V \\
e^{-2ik/3} W^*  & e^{- ik\/3} V^*  & -\Delta
\end{array} \right).
\eearr
This Hamiltonian is periodic in $6\pi $: $H (k+6\pi) = H(k)$,
due to the fact that the distance between the 
two successive sites within the unit cell is one-third of the
unit cell size.
Note that the canonical Bloch Hamiltonian $H(k)$ can be obtained from the
periodic Bloch Hamiltonian $\mathcal{H}(k)$ through the following unitary 
transformation: 
$ H(k) = U \mathcal{H}(k) U^\dagger$, where the unitary matrix $U$ is 
given by $ U= \text{diag}(1, e^{-ik/3}, e^{- i2k/3})$.

The canonical Bloch Hamiltonian $H(k)$ can be expressed in terms of the Gell-Mann matrices as
\bearr
H(k) 
& = & {\bs \Lambda} \cdot {\bm h(k)},
\eearr
where the components of the vector ${\bm h(k)}$ are 
$h_1 = u \cos(k/3) + u_1 \cos(2k/3), h_2 = u_1 \sin(2k/3)- u \sin(k/3), 
h_3 = \Delta/2, h_4 = w \cos(k/3) + w_1 \cos(2k/3),
h_5 = w \sin(k/3) - w_1 \sin(2k/3),
h_6 = v \cos(k/3) + v_1 \cos(2k/3), 
h_7 = v_1 \sin(2k/3) - v \sin(k/3)$, and
$h_8 = \sqrt{3} \Delta/2$.
The components $h_2,h_5,h_7$ are odd under time-reversal: 
$h_{2,5,7}(-k) = - h_{2,5,7}(k)$ and the remaining components are even under 
time-reversal.

\section{Different topological phases for $u_1=v_1\neq 0$ case} \label{Gen-bulk-boundary condition}
\begin{figure*}[htbp!]
\centering
 \includegraphics[width = 1\linewidth]{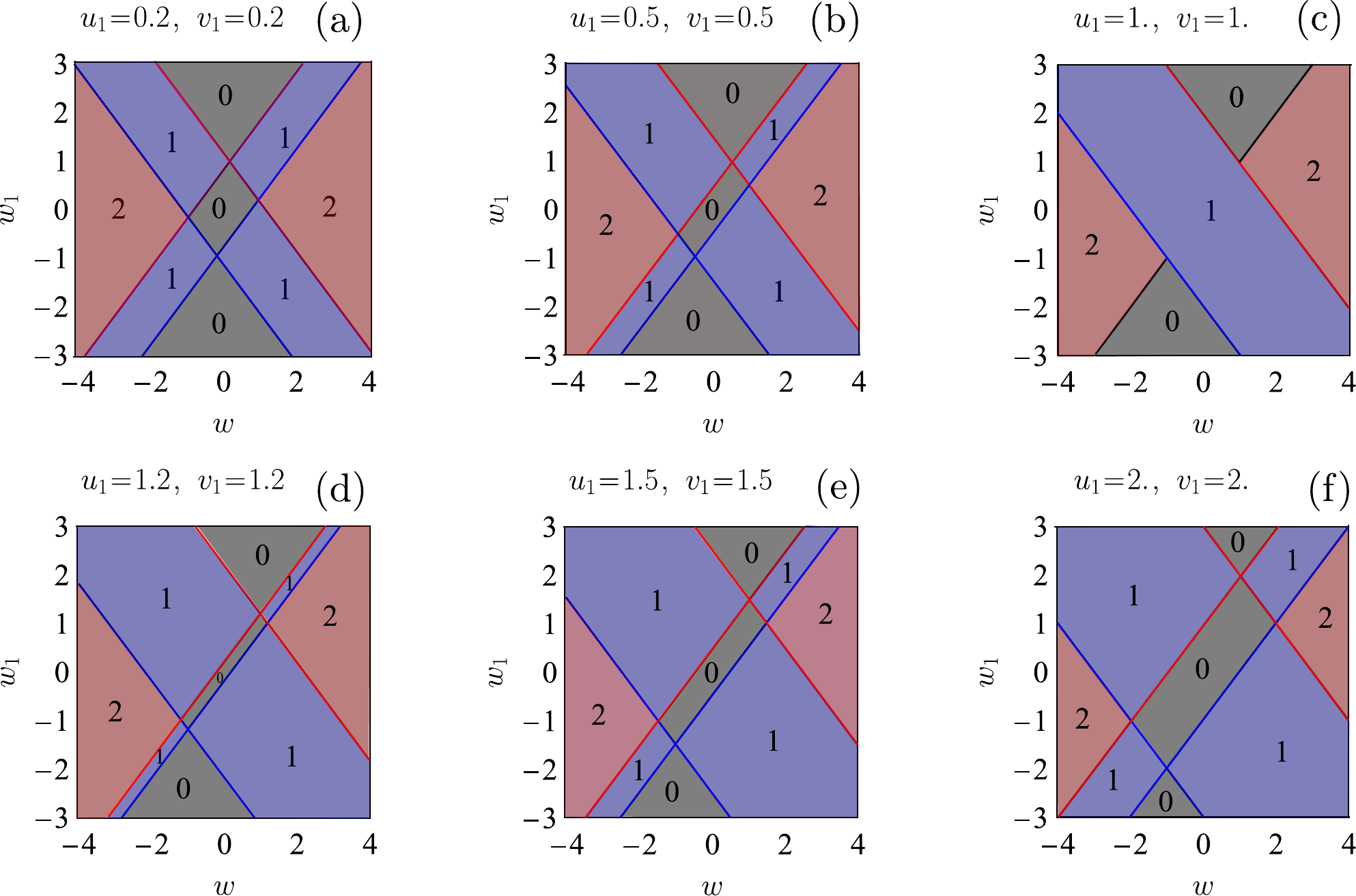}
\caption{\textbf{Bulk-topological invariant}: Here
(a), (b), (c), (d), (e), and (f) describe the sum of the Zak phases $\phi_1+\phi_3$
(in units of $\pi$) in the $(w, w_1)$ plane for 
$(u_1,v_1): (0.2,0.2), (0.5,0.5), (1.0,1.0), (1.2,1.2), (1.5,1.5)$, and $(2.0,2.0)$, respectively.
In all the plots here we have taken $u=v=1$.}
%
\label{fig:topological phase transitions23_Zak}
\end{figure*}

In Sec. \ref{sec:bulk-boundary condition}, we have presented results of different topological phases for $u_1=v_1=0$ case.
In this Appendix, we present results of different topological phases characterized by the Zak 
phases ($\phi_1 +\phi_3$) on $w$-$w_1$ parameter space for different sets of $u_1=v_1\neq 0$ case.  
The phase diagrams given in Fig.~\ref{fig:topological phase transitions23_Zak} look more complex and reveal that
the topologically non-trivial regimes corresponding to $\phi_1 +\phi_3 = 2\pi (\pi) $ decreases (increases) with the increase of the values of $u_1=v_1$.  

\counterwithin{figure}{section}


\begin{figure*}[htbp!]
\centering
 \includegraphics[width = 1\linewidth]{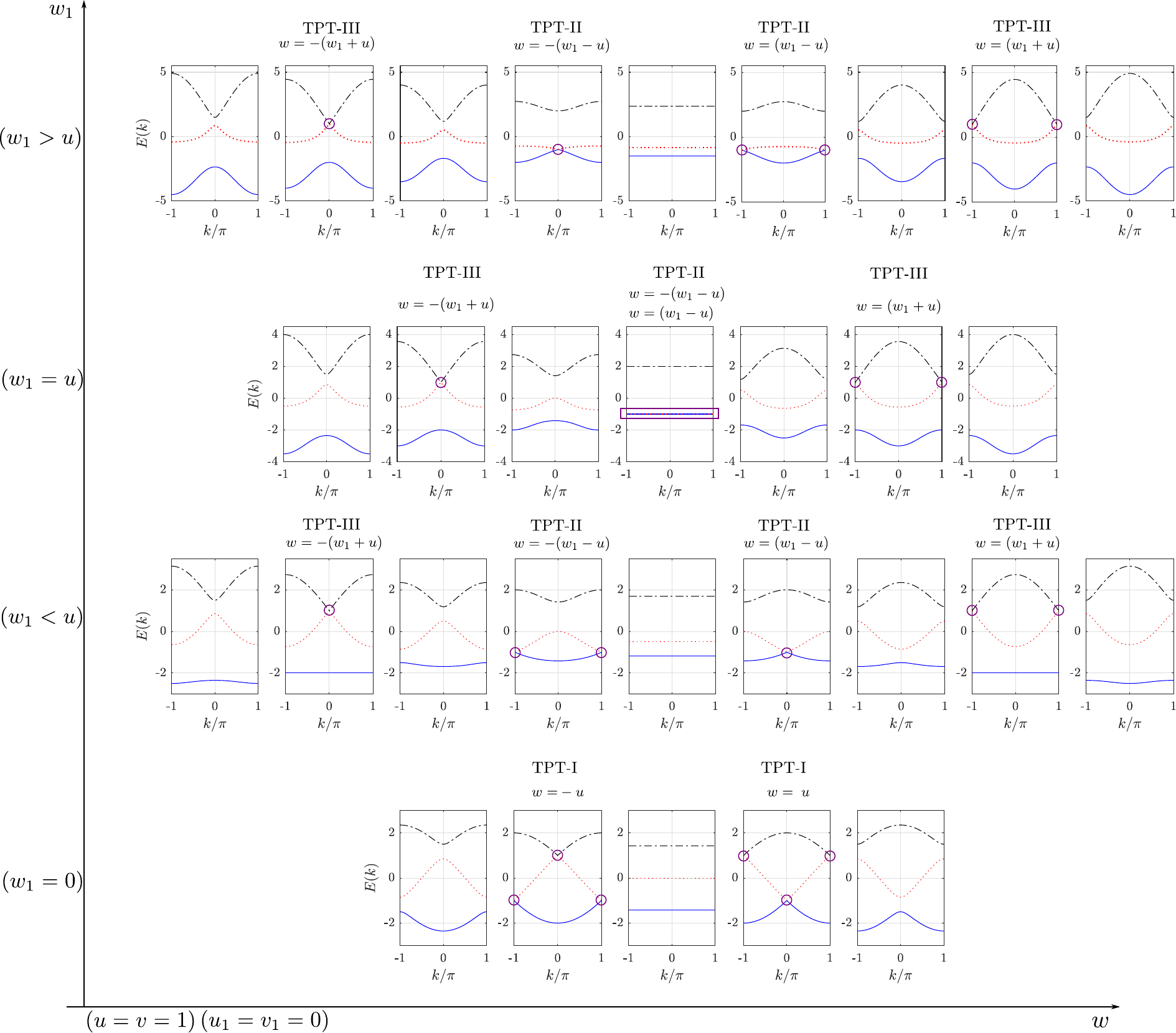}
\caption{\textbf{Bloch band structure}: 
In all the plots here, we have taken $u=v=1,~u_1=v_1=0$.}
%
\label{fig:topologicalphasetransitions123_band}
\end{figure*}

\section{Sub-lattice winding number for the dimer SSH model}\label{App:winding}
In this Appendix, we will present the results of the sub-lattice winding number for the dimer SSH model. 
The Bloch Hamiltonian of the dimer SSH model is given by 
\begin{align}
        h(k) & = 
        \begin{pmatrix}
        0 & d_k \\
        d_k^* & 0
        \end{pmatrix},
\end{align}
where $k$ denotes the Bloch wave vector and 
$ d_k = t_1 + t_2 e^{-i k}$, with $t_1$ and $t_2$ represent nearest-neighbour 
intra-cell and inter-cell hopping parameters, respectively. 
This Hamiltonian respects time-reversal, inversion, chiral, and particle-hole symmetries.
It supports boundary mode with energy $E_{\text b} = 0$ at each end of the system 
when $t_2> t_1$. 

It is straightforward to get the sub-lattice winding numbers for the 
sub-lattices $A$ and $B$ as
\begin{align}
\nu^{A}  & = \int_{-\pi}^{\pi} \partial_k \log(d_k) \frac{dk}{2\pi i} 
=\begin{cases}
       \; \;  0,~t_1>t_2\\
        -1,~t_1< t_2.
  \end{cases}
\end{align}
and
\begin{align}
\nu^{B}  &  = \int_{-\pi}^{\pi} \partial_k \log(d_k^*) \frac{dk}{2\pi i}
=\begin{cases}
       \; \;  0,~t_1>t_2\\
        +1,~t_1< t_2.
  \end{cases}.
\end{align}
The sub-lattice winding numbers $\nu^{A/B} $ are integer-valued and determine the
boundary modes at each end of the system. Thus the sub-lattice winding numbers
$\nu^{A/B}$ successfully specify the bulk-boundary
correspondence in the dimer SSH model as well.

\section{Real space profile of edge states and their connections to topological invariants}\label{App:real_edge}
Consider the eigenvalue equation in the open boundary condition,
\begin{align}
    H|\psi\rangle=\pm u |\psi\rangle,  
\end{align}
with 

$|\psi\rangle = (\psi_{1\rm A},\psi_{1\rm B},\psi_{1\rm C},\psi_{2\rm A},\psi_{2\rm B},\psi_{2\rm C},...,\psi_{N\rm A},\psi_{N\rm B},\psi_{N \rm C})^T$ and
\begin{align}
    H=\begin{pmatrix}
    0 & u & w_1 & 0 & 0 & 0 & 0 & 0 & 0\\
    u & 0 & v & u_1 & 0 & 0 & 0 & 0 & 0\\
    w_1 & v & 0 & w & v_1 & 0 & 0 & 0 & 0\\
    0 & u_1 & w & 0 & u & w_1 & 0 & 0 & 0\\
    0 & 0 & v_1 & u & 0 & v & u_1 & 0 & 0\\
    0 & 0 & 0 & w_1 & v & 0 & w & v_1 & 0\\
    0 & 0 & 0 & 0 & u_1 & w& 0 & u & w_1\\
    0 & 0 & 0 & 0 & 0 & v_1& u&0 &v\\
    0 & 0 & 0 & 0 & 0 & 0 & w_1 &v&0
    \end{pmatrix}
\end{align}
which immediately implies the following set of equations 
\begin{align}
   \mp u \psi_{1\rm A} + u\psi_{1\rm B} +w_1 \psi_{1\rm C}&=0,\\
   u \psi_{1\rm A} \mp u\psi_{1\rm B} +v \psi_{1\rm C} +u_1 \psi_{2\rm A}&=0,
   ...
\end{align}
For $u_1=0$, solving the above two equations immediately gives rise to \\
(i) $\psi_{1\rm C} =0, ~\psi_{1 \rm A}=\psi_{1\rm B}$ for $E_{\rm b}=-u$,\\
(ii) $\psi_{1\rm C} =0, ~\psi_{1 \rm A}=-\psi_{1\rm B}$ for $E_{\rm b}=u$.

Similarly, the eigenvalue equation for $E_{\rm b}=\pm v$ leads to 
\begin{align}
   w_1 \psi_{N\rm A} + v\psi_{N\rm B} \mp v \psi_{N\rm C}&=0,\\
   v_1 \psi_{(N-1)\rm C} + u\psi_{N\rm A} \mp v \psi_{N\rm B} +v \psi_{N\rm C}&=0,
   ...
\end{align}
For $v_1=0$, solving the above two equations immediately gives rise to \\
(i) $\psi_{N\rm A} =0, ~\psi_{N \rm B}=\psi_{N\rm C}$ for $E_{\rm b}=-v$,\\
(ii) $\psi_{N\rm A} =0, ~\psi_{N \rm B}=-\psi_{N\rm C}$ for $E_{\rm b}=v$.

Therefore, we conclude that the wave function for edge states with energies $E_{\rm b}=-u (u)$ and $E_{\rm b}=-v (v)$ have equal (equal and opposite) support on even and odd sub-lattice sites even in the presence of next-nearest neighbor hopping $w_1\neq 0$ where the point chiral and or inversion symmetries are lost.\\
\color{black}

\end{document}